\documentclass{aa}  
\usepackage[colorlinks=true,
    linkcolor=blue,
    filecolor=magenta,      
    urlcolor=cyan,
    citecolor=blue]{hyperref}

\usepackage{xcolor}
\usepackage{graphicx}
\usepackage{txfonts}
\usepackage{ulem}
\usepackage{multicol}
\usepackage{mathtools}
\usepackage{amsmath}
\usepackage[toc]{appendix}
\def\msun{{\rm M}_\odot}

\begin{document}

\title{Collision-induced mass loss and mass gain on an extremely massive star}

\subtitle{An analytical approach and a static proto-globular cluster test-case}

\authorrunning{Ramírez-Galeano et al.}
\titlerunning{Collision-induced mass loss on an aEMS in a proto-GC}

   \author{Laura Ramírez-Galeano
          \inst{1}, Corinne Charbonnel\inst{1,2}, Tassos Fragos \inst{1,3}, Zoubaïr Tazakkati\inst{4}, Jaime {Roman-Garza}\inst{1,3} and Mark Gieles\inst{5,6} }
          %\fnmsep\thanks{Just to show the usage
          %of the elements in the author field}}

   \institute{Department of Astronomy, University of Geneva, Chemin P\'egasi 51, 1290 Versoix, Switzerland
              \email{laura.ramirezgaleano@unige.ch}
         \and
            IRAP, UMR 5277 CNRS and Université de Toulouse, 14 Av. E.Belin, 31400 Toulouse, France
        \and 
            Gravitational Wave Science Center (GWSC), Universit\'e de Gen\`eve, 24 quai E. Ansermet, CH-1211 Geneva, Switzerland
        \and
            CMAP, CNRS, \'Ecole Polytechnique, Institut Polytechnique de Paris, 91120 Palaiseau, France
        \and            
            ICREA, Pg. Llu\'{i}s Companys 23, E08010 Barcelona, Spain
        \and
            Institut de Ci\`{e}ncies del Cosmos (ICCUB), Universitat de Barcelona (IEEC-UB), Mart\'{i} i Franqu\`{e}s 1, E08028 Barcelona, Spain
            }

   \date{Received; accepted:}

% \abstract{}{}{}{}{} 
% 5 {} token are mandatory
 
  \abstract
  {}
  {The objective of this study is to analytically explore mass loss and gain induced by stellar collisions on a gas-accreting extremely massive star (aEMS, $ 10^3 \lesssim 
 M/M_{\odot} \lesssim 10^4 $). We also consider  its contribution to the mass budget in the context of forming multiple stellar populations (MPs) in a typical protoglobular cluster (GC). } 
   {We used MESA to build a series of aEMS models up to $2 \times 10^4 M_{\odot}$ for three [Fe/H] values, covering the metallicity range of Galactic GCs, with different treatments of super-adiabatic convection. 
   We set analytical prescriptions to quantify collision-induced mass loss when a star spirals and deposits energy into the envelope of the aEMS. We used a Monte Carlo approach to simulate the effects of multiple collisions on an aEMS of initial mass of $10^3  M_{\odot}$ in a static proto-GC,  accounting for mass loss and gain from collisions, gas accretion, and stellar winds.}   
   {We show that assumptions on super-adiabaticity in radiation-dominated layers significantly impact the aEMSs properties and   their collision responses: extended stars tend to lose mass, while compact ones are more likely to gain it.  Our MC simulations predict the total mass lost and gained, along with the corresponding timescales and contributions from stellar winds and collisions. The results are influenced by the structural characteristics of the aEMS and by the gas accretion rate during the collision phase. Under certain conditions, the EMS exhibits a "conveyor belt" behavior, processing up to $10^{5.5} M_{\odot}$ of material in $\sim 5$~Myrs.}
   {This study provides theoretical predictions supporting aEMSs as contributors to the abundance anomalies observed in GCs.
   It emphasizes the importance of including collision dynamics and mass transfer in aEMS formation and evolution models in dense stellar environments. We provide a grid of predictions for stellar M-R-[Fe/H]-structure relations and collision-induced mass loss and gain, suitable for inclusion in hydro and N-body simulations of star clusters.}


   \maketitle
%
%-------------------------------------------------------------------

\section{Introduction}
\label{Introduction}

The formation and evolution of globular clusters (GCs) have long been subjects of interest for astrophysics and cosmology \citep{Krumholz2019}, with numerous questions 
that are bound to be resolved thanks to recent observational advances. 
A major breakthrough came over the past two decades from Hubble Space Telescope (HST) photometry combined with ground-based high-resolution spectroscopy on very large telescopes \citep[e.g.,][]{2004ApJ...605L.125B,Mucciarelli2009,2015AJ....149...91P,Dalessandro2016,2019MNRAS.487.3815M,2019MNRAS.486.5581G,2019MNRAS.483.1674R,2024arXiv240606683C}. This synergy revealed that all old Galactic and extragalactic GC scrutinized so far in the local universe host multiple stellar populations (MPs) with unique chemical properties. 
While their so-called first population (1P) stars share the chemical properties of halo stars of similar metallicity,  their second population (2P) stars exhibit diverse degrees of N, Na, Al, and He enhancements that are anticorrelated with O (and sometimes Mg) depletion (see \citealt{2019A&ARv..27....8G} for a review and \citealt{2023A&A...677A..73C} for more recent insight and references).  
The proportion of 2P stars in individual GC ranges between 30$\%$ and 80$\%$ and increases with the GC mass \citep{2006AJ....131.1766C,Carretta2010b,2017MNRAS.464.3636M,2019AJ....158..202L,2024arXiv240606683C}.  Similarly, the extent of the O-Na and Mg-Al anticorrelations increases with GC mass, with Mg depletion being observed only in the most metal-poor and/or most massive GC \citep[e.g.,][]{2017A&A...607A..44B,Pancino2017,2018ApJ...859...75M}. 
Helium variations between 1P and 2P are usually very small ($\Delta Y \simeq 0.01$), with the strongest ones ($\Delta Y \simeq 0.1$) derived in the most massive GC \citep[e.g.,][]{2018MNRAS.481.5098M,2023A&A...679L..13C}.

The new photometric window opened by the James Webb Space Telescope (JWST) on the coolest low-mass main sequence stars in GC has provided evidence that the same MP characteristics are found all over the color-magnitude diagram \citep{2022MNRAS.517..484N,2023MNRAS.522.2429M,2023ApJ...953...62Z,2024A&A...689A..59S}.
Similar MP signatures have also been detected in  extra-galactic massive star clusters with ages as low as $\sim 1.5$ Gyr \citep[e.g.,][]{2008AJ....136..375M,2020MNRAS.498.4472S,2020MNRAS.499.1200M,2022ApJ...927L..10K,2022MNRAS.515.2511S}.

There is a general agreement on the fact that 2P stars formed out of proto-GC gas mixed to various degrees (so-called dilution) with hydrogen-burning yields processed at high temperature and released in important quantities (so-called mass-budget problem) by a short-lived type of 1P so-called polluters \citep[][and references therein]{Prantzos2017}. Some debate persists however regarding the identity of the polluters \citep[e.g.,][and references therein]{Charbonnel2016,Bastian18,Milone2022}. However, we  know that 2P stars inherited only H-burning products, with very slight amount of He and no  products  of more advanced nucleosynthesis, and they are seen only in massive and compact star clusters, young and old, in addition to the mass budget issue \citep[e.g.,][]{2006A&A...458..135P,2011MNRAS.413.2297S}. This indicates that we are dealing with a very specific type of stellar polluters, which formed concurrently with their host proto-GC in very specific conditions and at any metallicity.

We followed the approach from \citet{Denissenkov2014}, who demonstrated that the dilution of the central H-burning products of a $10^4\,\msun$  super massive star (SMS) close to the zero age main sequence (ZAMS) with original proto-GC can reproduce the ubiquitous O-Na anticorrelation and the  scarce Mg-Al anticorrelation. \citet{Gieles+2018} proposed a scenario where SMS with masses above $10^4\,\msun$ can form at any redshift through runaway collisions in gas-rich and extremely compact cluster environments, independently of metallicity\footnote{This formation scenario differs from that of Pop~III SMS resulting from the direct collapse scenario in a metal-free environment \citep[e.g.,][]{2001Ap&SS.276..807O,2002Sci...295...93A,2002ApJ...564...23B,2007ApJ...654...66O,2018MNRAS.477.3694B,2023MNRAS.522.3256C}.}. 
The continuous rejuvenation of the SMS through collisions and gas accretion in \citet{Gieles+2018} model can provide enough H-burning ejecta to fulfil the requisite mass budget to account for the ratios between 1P and 2P stars. 

However, the required stellar density to form SMS in their runaway model is about four orders of magnitude higher than that inferred in the proto-GCs that have been identified in young and confirmed bound systems in gravitationally lensed galaxies at redshifts up to 10.6 - 11 \citep{2024arXiv240103224A}. In addition, \citet{Gieles+2018} assumed total mass conservation regarding the products of the collisions. This is the hypothesis we are aiming to test in this paper.   

On the other hand, \citet{Prantzos2017} showed that protocluster pollution by early main sequence stars with masses between $ \sim 10^3$ and $10^4\, \msun$ (which they also call SMS, and with actual values depending on metallicity) can also explain the abundance properties of MPs. According to the inertial inflow \citep{Padoan+2020} GC formation model by \citet{Gieles+2024}, stars in this mass range could  form in proto-GCs (with masses of $\gtrsim10^5\,\msun$) via gas accretion as a natural extension of the IMF beyond very massive stars (VMS, with masses between $\sim 10^{2}$ and $10^{3}$ M$_{\odot}$), whose existence has been established in extragalactic super star clusters \citep[e.g.,][]{2016MNRAS.458..624C,2016ApJ...823...38S,2023ApJ...958..194S,2023A&A...673A..50M,2023A&A...679A..36S,2024A&A...686A.185U}.
Here, we follow the nomenclature of \citet{Gieles+2024}, who called such objects accreting extremely massive stars (aEMS; $ 10^{3-4}\, \msun$). In their model, aEMS can reach an equilibrium between gas accretion and wind mass loss. This conveyor belt behavior is expected to rejuvenate the aEMS by supplying fresh H, making the star stay in the MS phase as long as the conveyor belt phase, thereby allowing it to live longer than what would traditionally be expected. Consequently, it releases H-burning yields with low He contamination and in sufficient quantities regarding the mass budget issue. However, stellar collisions, which cannot be avoided in dense stellar systems, might also interfere during the growth of VMS and aEMS, potentially leading to an intermediate picture between the runaway SMS and the aEMS scenarios to explain the origin of MPs in GCs.  

The interest in collisions in star clusters also follows a long history. A number of theoretical and numerical studies have demonstrated that under specific conditions, stellar collisions and mergers can potentially result in the formation of VMSs, SMSs, and eventually IMBHs \citep[][]{1970ApJ...162..791S,1989ApJ...343..725Q,2002ApJ...576..899P,2004ApJ...604..632G,2006MNRAS.368..141F,2015MNRAS.454.3150G,2016MNRAS.459.3432M,2024ApJ...969...29G,vergara2025rapidformationmassivestar}. However, the collision path and its end product have been demonstrated to be influenced by several factors \citep[e.g.,][]{ Glebbeek+2009,2024ApJ...969...29G}. The initial conditions of the cluster (mainly its mass and radius) represent the primary parameter that might drive its early and long-term dynamical evolution. These can be constrained by the type of observations described above or derived from galactic simulations in a cosmological context with the requisite resolution to allow the formation of bound clusters to be followed in a robust manner \citep[e.g.,][]{2020IAUS..351...34L,2023MNRAS.521..124R,2024A&A...686A.106B,2024MNRAS.530.2760D,pascale2025siegeivcompactstar}. 
Other parameters are important, such as the stellar density and initial mass function (IMF), mass segregation, and the initial fraction of binaries.

The model predictions are also affected by the treatment of the evolution of single and binary stars (mass-radius relation, lifetime, ...) as well as the assumptions about the 
resulting merger objects (mass, radius, rejuvenation, remnant), which are crucial, yet highly uncertain ingredients. Smoothed particle hydrodynamical simulations have been employed to model the collision of two stars, initially focusing on low-mass stars in the context of exploring the formation of blue stragglers in star clusters \citep[e.g.,][]{2002ApJ...568..939L}, but also extending to more massive stars \citep[e.g.,][]{2008MNRAS.383L...5G}, which are the potential progenitors of neutron stars and black holes. Further evolving a merger product is numerically challenging, as it involves mapping back the still not fully relaxed stellar mass distribution from the three-dimensional (3D) hydrodynamic simulation to a one-dimensional (1D) stellar evolution code  \citep[e.g.,][]{1997ApJ...487..290S,2001ApJ...548..323S,2013MNRAS.434.3497G,2016MNRAS.457.2355S, 2019Natur.574..211S,2020MNRAS.495.2796S,2020ApJ...901...44W}. 
All the aforementioned studies look at stellar collisions with mass ratio in the range of $\simeq 1-10$ and do not consider stellar masses in the VMS+EMS+SMS regime. 
Therefore, the hydro and N-body simulations of massive and compact star clusters lack necessary and tractable input for mass-loss induced by collisions.
Furthermore, while some of these models follow the dynamical evolution of dense star clusters over extended periods, they do not address the early phases of the evolution of their constituent stars before they arrive on the ZAMS.
Given that pre-main sequence  stars have larger radii and different internal structures and chemical compositions than their main sequence descendants, this could have implications for the collision rates, the collision-induced mass loss, and the merger product, as well as the maximum mass a star could reach in a cluster. 

The present study aims to provide predictions for collision-induced mass loss using an analytical approach based on the mass-dependent structure of aEMSs that allowed us to cover a broad range of stellar masses and radii for both the aEMS targets and the colliding stars in the very early phases of cluster formation. 
In Sect. \ref{makingthemodels}, we describe how we built the aEMS models that are required for our analysis; we discuss how the uncertainties in the transport of energy in their convective layers affect the predictions for their global properties and compare with models from the literature. The nucleosynthesis behavior of aEMSs is presented in Appendix \ref{nucleosynthesis-appendix}. 
In Sect.\ref{onecollision}, we set the analytical prescriptions we then used to quantify collision-induced mass loss and gain. We identified the main processes at work, considering inspiraling and target stars over a broad mass range. We provide grids with model predictions that can be used for N-body and/or hydro simulations of star clusters.
As a first example of application, in Sect. \ref{multiplecollision}, 
we detail our  use of a Monte Carlo approach to simulate multiple collisions on an aEMS target in a dense static proto-cluster during its first 5~Myrs and we derived the amount of mass it can lose and gain considering collisions, gas accretion, and mass loss via stellar winds. 
We summarize our results and the sources of uncertainties of our predictions in Sect. \ref{discussion}.

%--------------------------------------------------------------------
\section{Mass-dependent structure of aEMS models} 
\label{makingthemodels}

\subsection{Preparatory aEMS model calculations}
\label{InputPhysics}

Our purpose is to create mass-dependent structures of aEMS to evaluate collision-induced mass loss. We used the 1D stellar evolution code Modules for Experiments in Stellar Astrophysics \citep[MESA][]{Paxton2011, Paxton2013, Paxton2015, Paxton+2018, Paxton2019, Jermyn_2023}\footnote{We use MESA version 22.11.1 together with the 23.7.2 version of the MESA software development kit.}.
Since in most GCs 2P stars are hardly enriched in He with respect to 1P stars, aEMSs are expected to acquire mass and eject the nucleosynthesis products of hot H-burning very early on in the main sequence. 
Accordingly, we assumed a fast gas accretion path, starting from a 0.7~$\msun$ seed,   until the models reach the desired mass range.  We  opted for the accreting formulation of \citet[their Eq.~1, 2, and 3]{Haemmerle+2019}, which is based on the Churchwell-Henning (CH) empirical relation \citep{Churchwell1999,Henning2000} and scaled it up by a factor of 10.  In practice, this implies a mass-dependent mass-accretion rate that increases from  $\sim 10^{-4}$M$_{\odot}$yr$^{-1}$ (with the advantage of avoiding numerical instabilities at the beginning of the calculations) to $ 6.15 \times 10^{-2}$ (respectively $\sim 10^{-1}$)  M$_{\odot}$yr$^{-1}$ when the model reaches  $10^3$ (respectively  $10^4$) M$_{\odot}$, with a mean value of a $\sim 4 \times 10^{-2}$ M$_{\odot}$yr$^{-1}$. These values are of the right order of magnitude compare to the aEMS model of \citet{Gieles+2024}. 
Mass loss through stellar winds is not included in these preparatory calculations, as the numerical objective was to ensure a fast total mass growth rate of the models while limiting the He-enrichment when the models enter the EMS regime. This is described in Appendix \ref{nucleosynthesis-appendix} where we provide more information about the input microphysics of the models and discuss their nucleosynthesis behavior. We also explore the effects of metallicity on the internal structure and nucleosynthesis of aEMS by computing models for three different metallicity values corresponding to well-characterized Galactic GCs ([Fe/H] = -0.72, -1.14, and -2.00 dex, for NGC 104 (47Tuc), NGC 2808, and NGC 6397, respectively).

Since the mass-dependent structure of aEMS is 
what matters for our collision study, it is important to understand which assumptions made in the preparatory calculations affect the stellar properties. The impact of the gas accretion rate has been extensively studied in the literature \citep[e.g.,][]{Hosokawa_2013,Herrington+2023,2018MNRAS.474.2757H,2024A&A...689A.351N} and is recalled in brief below. However (and as discussed in Sect.~\ref{Convection}) for the first time, the treatment of super-adiabatic convection has an important impact on the aEMS properties even for high accretion rates.

\subsection{Implications of the treatment of super-adiabatic convection in radiation-dominated stellar layers on the aEMS global properties}
\label{Convection}

\begin{figure*}[ht!]
    \centering
    \includegraphics[width=\textwidth]{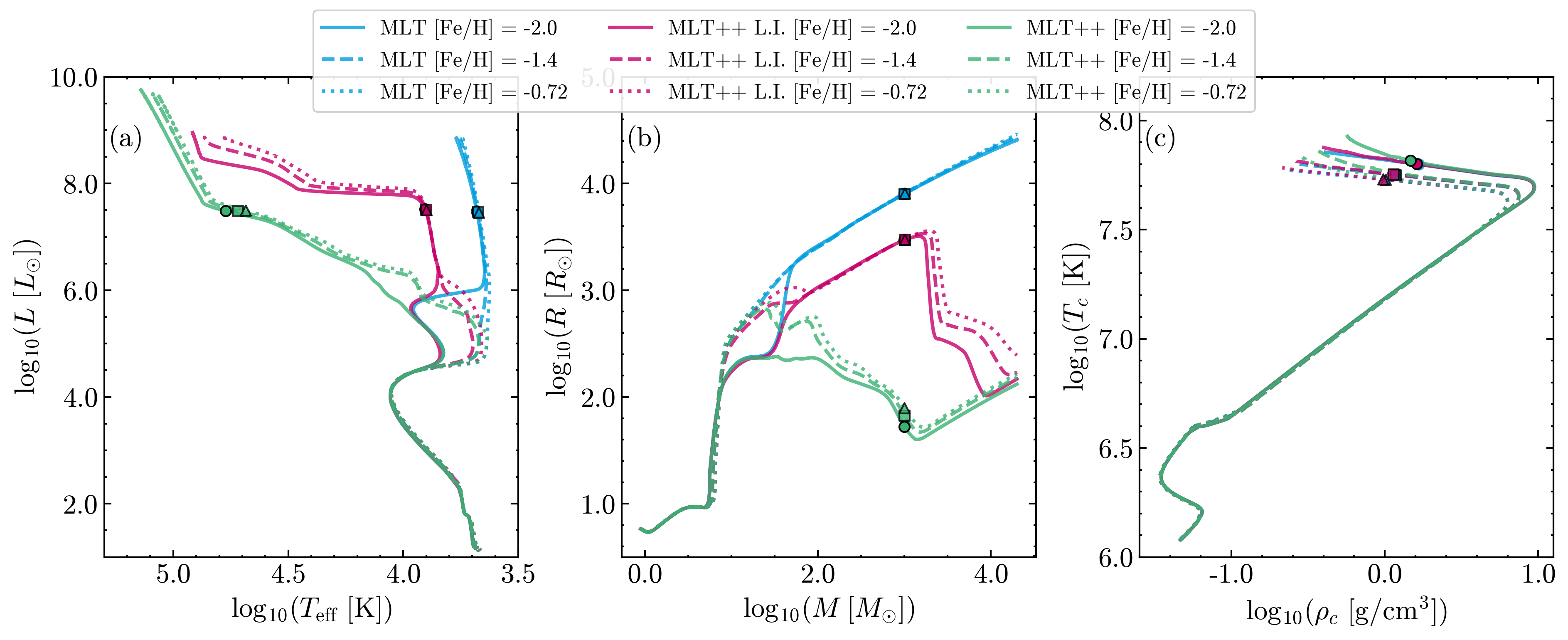} 
    \caption{Evolution of the key properties of the accreting models starting from $0.7 M_{\odot}$ and reaching $ 2 \times 10^4 M_{\odot}$ (the circles, squares and triangles indicate the points at which the accreting models reach a mass of $1000 M_{\odot}$ for reference). The models are computed with the same accretion rate and no mass loss, for the reasons discussed in the text.} \label{fig:hrd}
\end{figure*}

\begin{figure*}
    \centering
    \includegraphics[width=\textwidth]{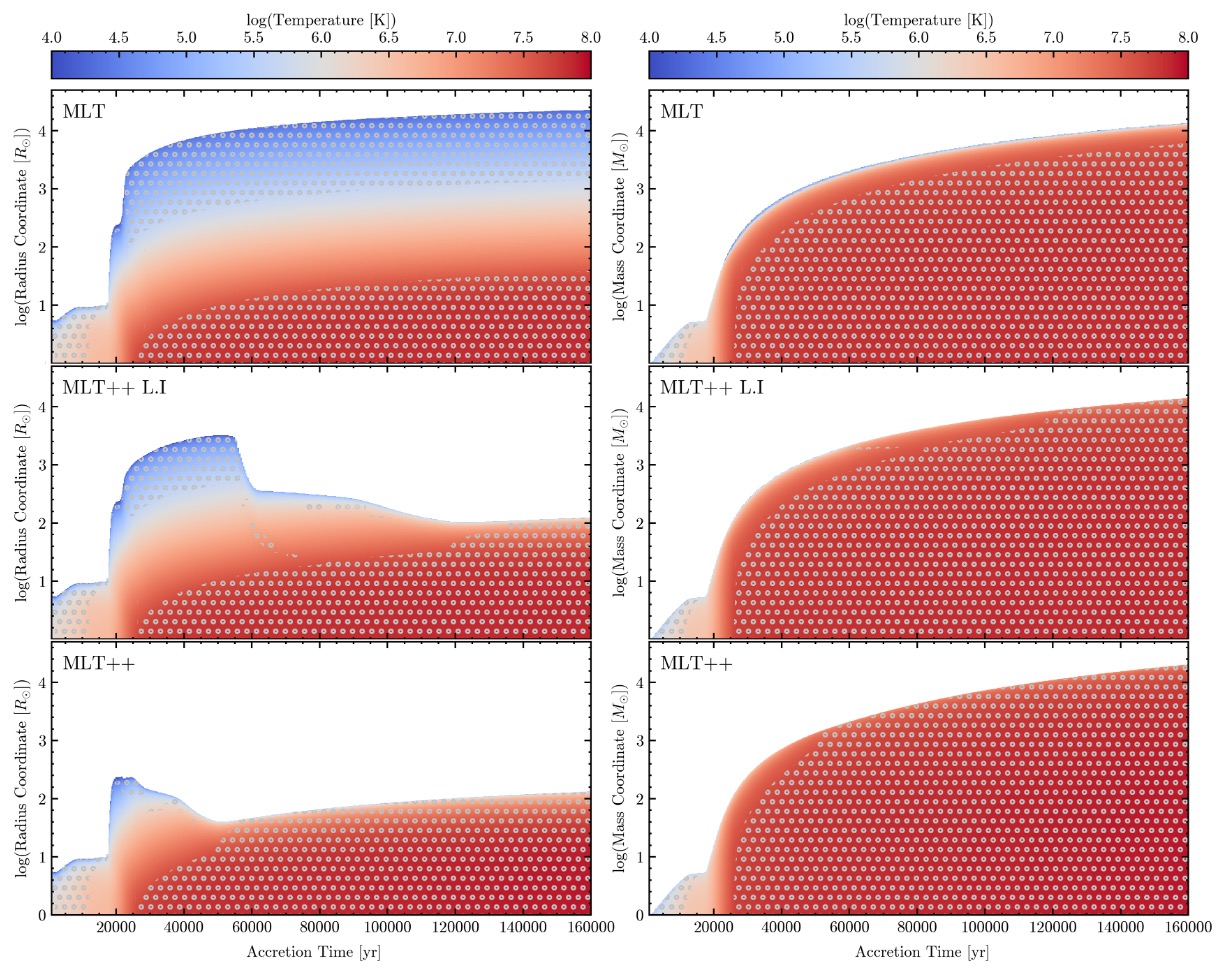} 
    \caption{Kippenhahn diagrams in radius and mass coordinates (left and right respectively) showing the evolution of the internal structure of the accreting models with [Fe/H]=-2~dex as in Fig.~\ref{fig:hrd} for MLT, MLT++L.I., and MLT++ (from top to bottom). 
    The x-axis is the time since the beginning of accretion on the 0.7~$\msun$ seed. The color maps show the temperature profiles. Areas highlighted with gray dots denote convective regions.} 
    \label{fig:kipp}
\end{figure*}

The efficiency of energy transport by convection in massive stars has long been debated \citep[e.g.,][]{1997ASPC..120...83L,2009pfer.book.....M}. More recently, 3D radiative hydrodynamic simulations have begun to provide insights into the intricate physical processes occurring in this regime. These include the effects of hydrodynamic instabilities and of waves excited by near-sonic turbulent convection \citep[e.g.,][]{2015ApJ...813...74J,2023Galax..11..105J,2020ApJ...902...67S}. Such phenomena cannot be adequately addressed within the confines of 1D stellar evolution models. This leads to very uncertain stellar radii in proximity to or above the Eddington limit, where radiative forces significantly influence convective flows and the overall energy transport mechanism \citep[]{Maeder+2008,2013MNRAS.433.1114Y,Kohler+2015,Sanyal+2015,2021A&A...647A..13G,Agrawal+2022,2022MNRAS.512.5717A}. 

The challenges described above also apply to aEMSs. These radiation-pressure-dominated objects exhibit super-Eddington luminosities with values between approximately $10^8$ and $10^{10}$ $L_{\odot}$, as shown in Fig.~\ref{fig:hrd}.  There we present key properties of the accreting models with the three values of [Fe/H] that we computed with two different formalisms for the transport of energy by convection in addition to the classical MLT \citep{Mihalas1978, Paxton2011}.
The so-called MLT++ formalism was developed by \citet[][]{Paxton2013} to artificially reduce the super-adiabaticity implied by MLT in some radiation-dominated convective regions of massive stars nearing the Eddington limit. It uses nonlocal modifications to the energy transport equations to better handle some numerical stability issues that arise near the Eddington limit. 
A modified version of MLT++, here referred to as MLT++L.I. \citep[][]{Jermyn_2023}, also artificially enhances the transport of energy by convection when the Eddington factor is high, but incorporates a fully implicit and local formulation. It uses automatic differentiation to improve the stability and performance of the calculations across a wide range of masses and metallicities. As clearly stated by \citet[][]{Jermyn_2023}, both MLT++ and MLT++L.I. are meant to serve as "stellar-engineering methods to circumvent complex evolutionary stages, rather than a specific physical model that accounts for how convection is modified near the Eddington limit."Nevertheless, the comparison between the corresponding models that we present below allows for the investigation of the range of uncertainty on the stellar radius and internal structure of aEMSs due to our lack of knowledge about the transport of energy by convection in this regime.  This is of particular importance, as it will impact the outputs of the collision-induced processes that we study. 

Panel (b) in Fig.~\ref{fig:hrd} illustrates the monotonic increase in the radius of the MLT model until the aEMS reaches the imposed maximum mass of $2 \times 10^4 M_\odot$. The gas accretion rate is such that the advected entropy cannot be quickly radiated away, and the Kelvin-Helmholtz contraction timescale of the stellar interior is always too small compared to the increase in stellar radius induced by accretion, independently on the metallicity.  Consequently, for the three [Fe/H] values we consider the MLT models are predicted to climb the Hayashi track as supergiant protostars that are highly inflated and red (see the blue tracks in the Hertzsprung-Russell diagram (HRD) in panel (a) of Fig.~\ref{fig:hrd}). These findings obtained with the MLT framework are in excellent agreement with models with high accretion rates typically above $10^{-2}$ M$_{\odot}$yr$^{-1}$ (as in our models) from the literature \citep[][]{Hosokawa_2013,Herrington+2023,2018MNRAS.474.2757H,2024A&A...689A.351N}. The treatment of convection in these studies is not specified, which may be indicative of the similarity of their assumptions to those of our MLT models.

The radius of the models computed with the two other convection treatments undergoes inflation, similarly to what is seen  for the MLT model during the first $\sim 30$Kyr (MLT++) and $\sim 50$Kyr (MLT++ L.I.), reaching up to a radius of $\sim 2400R_{\odot}$ and $\sim 3000 R_{\odot}$, respectively (these numbers are for [Fe/H]=-2~dex and slightly depend on metallicity). 
However, and as expected, there are substantial differences in the radius and effective temperature ($T_{\text{eff}}$) of the models, depending on the convection treatments in the regime where radiative pressure is significant. For the [Fe/H]=-2~dex case, this happens for log(L/L$_{\odot}$) above $\sim$ 5.8, which is reached when the model has a mass of $\sim$ 40 M$_{\odot}$. Due to the imposed increase in the efficiency of the energy transport by convection, the MLT++ and MLT++L.I. models manage to evacuate the excess energy resulting from the gravitational settling of the accreted material after approximately one thermal time, causing them to deflate to smaller radii despite the relatively high accretion rates we assume. Whatever the metallicity, the MLT++ model is the most compact one, with the highest $T_{\mathrm{eff}}$, due to the artificial enhancement of the energy transport from the core to the surface. The MLT++L.I. formalism maintains an entropy gradient that is sufficient to drive convection effectively, but not so steep as to cause excessive compactness or elevated surface temperatures. This results in a model that is intermediate in compactness and $T_{\mathrm{eff}}$  between MLT and MLT++.

The uncertainty on the stellar radius in the aEMS regime due to our lack of knowledge on the energy transport efficiency by convection, it will result in some uncertainty in the amount of mass that can be lost by the collision of a star with the aEMS, as discussed in Sections~\ref{onecollision} and \ref{multiplecollision}.  

\subsection{Implications of the treatment of super-adiabatic convection in radiation-dominated stellar layers on the  internal structure and central conditions of aEMS}
\label{Convection-centraleffects}

The Kippenhahn diagrams in radius and mass shown in Fig.~\ref{fig:kipp} for [Fe/H]=-2~dex illustrate the evolution of the internal structure depending on the convection treatment. 
Upon entering the EMS regime (around $25\times 10^3$ years) with continuous expansion of the stellar radius, the MLT model permanently exhibits two convective regions separated by an extended (in radius) radiative zone. The outer convective region is a consequence of the high opacity in most external layers of this model, which has the lowest $T_\mathrm{{eff}}$ and the largest radius. The convective core is driven by H-burning at high temperatures.  
A large part of the interior of this model is not thermally relaxed, with a positive outward entropy gradient created by gas accretion that prevents the expansion of the convective core. More specifically, since the Brunt-Väisälä N$^2$ frequency is directly proportional to the entropy gradient ds/dr, the expansion in radius of the convective core is quenched. 
While the external and internal convective regions have similar extensions in radius coordinates, the outer region corresponds to only about $\sim6\%$ of the star's total mass, whereas the inner region corresponds to about $\sim44\%$.
The MLT++ L.I. model initially resembles the MLT model, featuring two convective zones until approximately $6\times 10^4$ years when the mass is around $2\times 10^{3} M_{\odot} $. Subsequently, when the star contracts, the effective temperature and the temperature of the outer layers increase, leading to a decrease in the opacity and convection in the outer layers. However, as the evolution proceeds, the convective core expands both in radius and mass, and the star becomes fully convective once it has reached a mass of $\sim$ $10^4$$M_{\odot}$ around 135 kyrs.  In the extreme MLT++ case, the star is even more compact and becomes fully convective after $54$ kyrs when its mass has reached $\sim$ $1.7 \times 10^3 M_{\odot}$. 
In all cases, more than $30\%$ of the mass of the aEMS is convective. 

Concerning the central conditions, the models with the three different convection treatments at a given metallicity follow similar tracks in the $T_c-\rho_c$ plot (Fig.~\ref{fig:hrd}, panel (c)), with the more compact MLT++ model reaching slightly higher central temperatures than the other two. 
This shows that while the convection treatment affects the macroscopic properties of the models, it does not significantly influence the conditions at their center nor the nucleosynthesis at a given metallicity. We refer to Appendix \ref{nucleosynthesis-appendix} where we show that the aEMS mass range ($10^3 - 10^4 \msun $) provides the relevant nucleosynthesis to explain MSP properties \citep[see also][]{Gieles+2024}.

\begin{figure*}
    \centering
    \includegraphics[width=\textwidth]{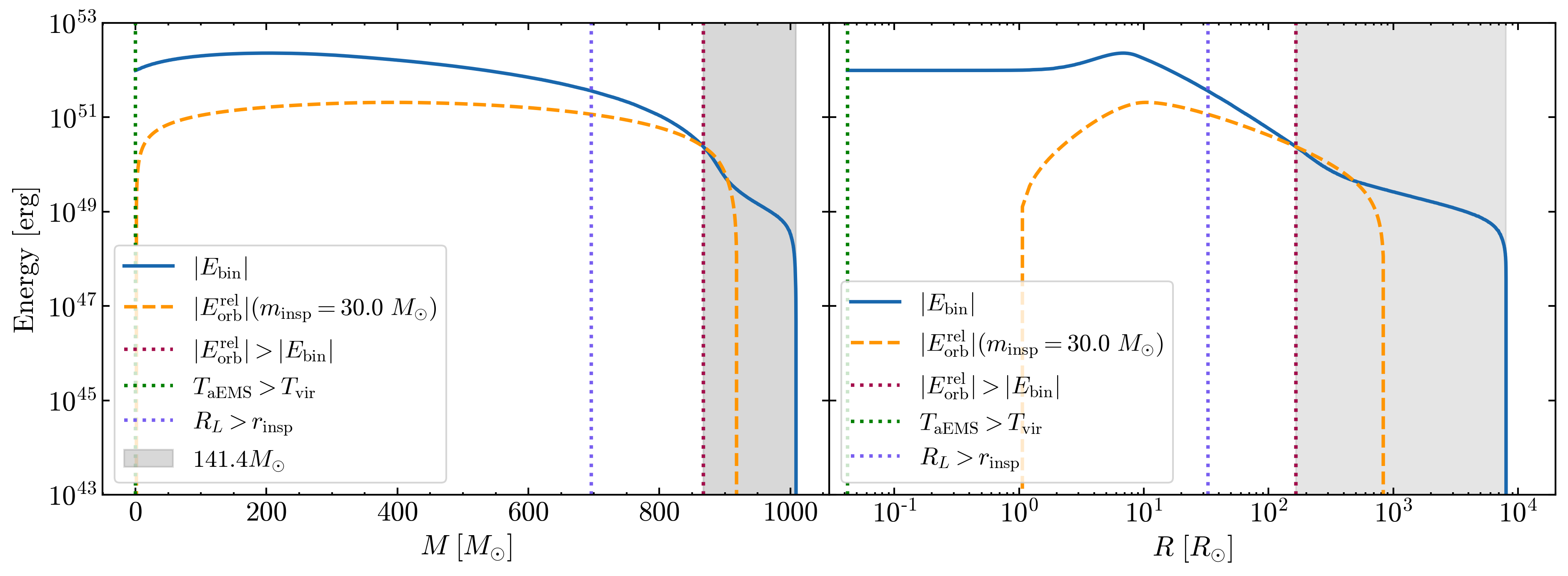}
    \caption{Comparison between  $|E_\mathrm{bin}|$ within an aEMS of $1000M_{\odot}$  and the orbital energy $|E_\mathrm{orb}^\mathrm{rel}|$ of 
    a $30M_{\odot}$ inspiraling star (blue full  and orange dotted lines respectively), for models computed with MLT and [Fe/H]=-2.00~dex,
    with the abscissa in mass and radius in the left and right panels, respectively.
    The red-dotted lines show the coordinates where $|E_\mathrm{orb}^\mathrm{rel}|$ is higher than $|E_\mathrm{bin}|$. The green-dotted lines represent the points where the local temperature of the aEMS exceeds the virial temperature of the inspiraling star. The dotted-purple lines represent the points where the radius of the inspiraling star fills the Roche lobe.}
    \label{fig:energy_mltpp}
\end{figure*}

\subsection{Other sources of uncertainties}
\label{others}

Beyond the uncertainties regarding convection, an additional  source for the transport of energy and entropy in the  radiative layers of the aEMS could be turbulence induced by fast accretion shocks and/or collisions. The 3D radiative-hydrodynamical simulations of the second gravitational collapse following the dissociation of H$_2$ molecules  \citep{1969MNRAS.145..271L} and leading to the birth of a low-mass protostar ($1~M_{\odot}$) have shown that strong turbulence develops inside their radiative interior and is maintained as long as accretion precedes onto the protostar 
(\citealt{2023A&A...680A..23A}; see also the 2D simulations from \citealt{2020A&A...638A..86B} for protostars up to $20 M_{\odot}$).   
Whether or not such a strong turbulence could be triggered and maintained in the aEMS remains to be investigated, along with its efficiency in transporting entropy that could potentially  moderate the swelling of aEMS. We assume that this phenomena has no impact on our conclusions regarding collision-induced mass loss discussed in the next sections, where we probe the impact of the convection treatment that is the main driver of the swelling efficiency. We note, however, that the presence and efficiency of turbulence in the radiative layers could transport the H-burning products from the core to the external layers, allowing their ejection in the wind, and/or during the peeling of the aEMS due to stellar collisions.

\section{The impact of a single collision - Analytical approach}
\label{onecollision}

Here, we focus on collision-induced mass loss when a star (hereafter inspiraling star) spirals and deposits orbital energy into the envelope of an aEMS (hereafter target star). 
We studied the effects of this energy deposition inside the aEMS analytically (i.e., we do not follow the hydrodynamical reaction of the aEMS to the collisions), following the physical criteria described in Sect.~\ref{Analytical_prescriptions}.

In Sect.~\ref{Analytical_predictions}, we explain how we used the internal profiles of the relevant physical quantities from the aEMS  internal structures of the fast gas accreting models at different masses to infer whether a collision will result in mass loss from the aEMS. We considered three cases for the treatment of convection, but focus the discussion on the models with [Fe/H]=-2.00~dex.

\subsection{Analytical prescriptions}
\label{Analytical_prescriptions}

We assumed as an initial condition that the inspiraling star is in a circular orbit just below the surface of the target aEMS. Then, as a result of the drag that it feels from the surrounding gas, it starts inspiraling within the envelope of the aEMS. Given the gas properties of an aEMS's envelope and the physical size of a typical star, the dominant drag is expected to be the hydrodynamic one, which is the only drag mechanism we consider. 
We expect the inspiraling star of mass $m_\mathrm{insp}$ and radius $r_\mathrm{insp}$ to continue its inward journey as long as two conditions are met: (1) its virial temperature \citep{Siess+1999,Privitera+2016}, 

\begin{equation}
    T_\mathrm{virial} = \frac{\mu  m_\mathrm{H}  G  m_\mathrm{insp} }{k_B  r_\mathrm{insp}},
    \label{eq:tvir}
\end{equation}
remains higher than the local temperature $T_{\text{aEMS}}(r)$ within the aEMS; then
(2) its radius remains smaller than its Roche lobe radius, which we compute as \citep{Eggleton83}:

\begin{equation}
    R_{L} = r \frac{0.49\ q^{2/3}}{0.6\ q^{2/3}+ \ln (1+q^{1/3})},\  q = m_\mathrm{insp}/M_\mathrm{aEMS}(r)
    \label{eq:roche_lobe}
,\end{equation}
with $r$ being the orbital separation of the system, namely, the distance between the center of the target aEMS and the position of the inspiraling in its envelope ($r \leq R_\mathrm{{aEMS}}$, as 
we assume that the inspiraling star is in an orbit below the surface of the target aEMS).
If either of these conditions is not met at a given depth within the aEMS, we assume that the inspiraling star has merged with the aEMS, transferring all of its mass to the latter. 
As for the mass-radius relations, we consider the model predictions for the three treatments of convection for both the inspiraling and the target stars. However, for the inspiraling stars,  we did not apply the factor of 10 to the  CH relation.This is in agreement with the maximum accretion rates for stars with final masses of $\sim 10^{2} \msun$ from the inertial flow simulations of \citet{Padoan+2020}.

The orbital energy that is lost is released locally in the form of heat within the aEMS's envelope. 
Since we  performed our calculations in 1D, we made the implicit assumption that this energy is instantaneously distributed within a spherical shell of the aEMS, around the location of the inspiraling star. If the energy released by the inspiraling star during its orbit is greater than the binding energy of the aEMS's layers above the current position of the orbit, as well as the energy transport timescale from that point to the surface (as reported by our stellar models) is longer than the inspiraling time scale,
it can cause the aEMS to lose mass. 
The orbital energy of the inspiraling star at a given distance $r$ from the center of the aEMS
is computed as
\begin{equation}
    E_\mathrm{orb}(r) = -\frac{GM_\mathrm{aEMS}(r)m_\mathrm{insp}}{2r},
    \label{eq:Eorb}
\end{equation}
where $M_\mathrm{aEMS}(r)$ is the enclosed mass of the aEMS at $r$. 
The amount of orbital energy released is quantified as $E_\mathrm{orb}^\mathrm{rel} = E_\mathrm{orb}^\mathrm{ini}  - E_\mathrm{orb}$, with $E_\mathrm{orb}^\mathrm{ini}$ the orbital energy at the surface of the aEMS. 
The binding energy of the aEMS, denoted as $E_\mathrm{bind}$, is determined as the addition of the potential energy, 
\begin{equation}
    U = - G \int^{M_\text{aEMS}(r)}_{0} \frac{m'}{r'}dm'
    \label{eq:Epot}
,\end{equation}
and its internal energy,  $I$, which is one of the structure outputs of the models.

As mentioned above, both the amount of orbital energy deposited and the corresponding timescale, $\tau_{dr} = r/\dot{r}$, are important.
If $\tau_{dr}$ is longer than the thermal timescale $\tau_{th}$, we assume that no mass loss would be expected from the aEMS because of the injected energy can be dissipated efficiently, even in the cases where the energy deposited by the inspiraling star exceeds the binding energy of the aEMS's envelope layers outside the orbit.
To compute $\tau_{dr}$, we  ought to consider $\dot{r}$ as proportional to the rate of orbital energy decay attributed to the drag, which we express as  

\begin{equation}
    \dot{r} = \dot{E}_{d} \times \left[ - \frac{1}{2} \frac{Gm_\text{insp}}{r}\left(\frac{dM_\text{aEMS}}{dr} - \frac{M_\text{aEMS}(r)}{r}\right) \right]^{-1},
    \label{eq:rdot}
\end{equation}
where $\dot{E}_{\text{d}} \approx F_{\text{d},x}v_\mathrm{cir}$. $F_{\text{d},x}$ represents the drag force experienced by the inspiraling star along the direction of its orbital motion. This can be written as

\begin{equation}
    F_{\text{d},x} = \pi r^2_{\text{insp}}\rho v_\mathrm{cir}^2,
    \label{eq:Fdx},
\end{equation}
where $v_\mathrm{cir}$ denotes the circular velocity of the orbiting star at that radius, namely, 

\begin{equation}
    v_\mathrm{cir}= \left[\frac{G (M_{\text{aEMS}}(r) + m_{\text{insp}})}{r}\right]^{1/2}.
    \label{eq:Vcir}
\end{equation}

\begin{figure*}
    \centering
    \includegraphics[width=\textwidth]{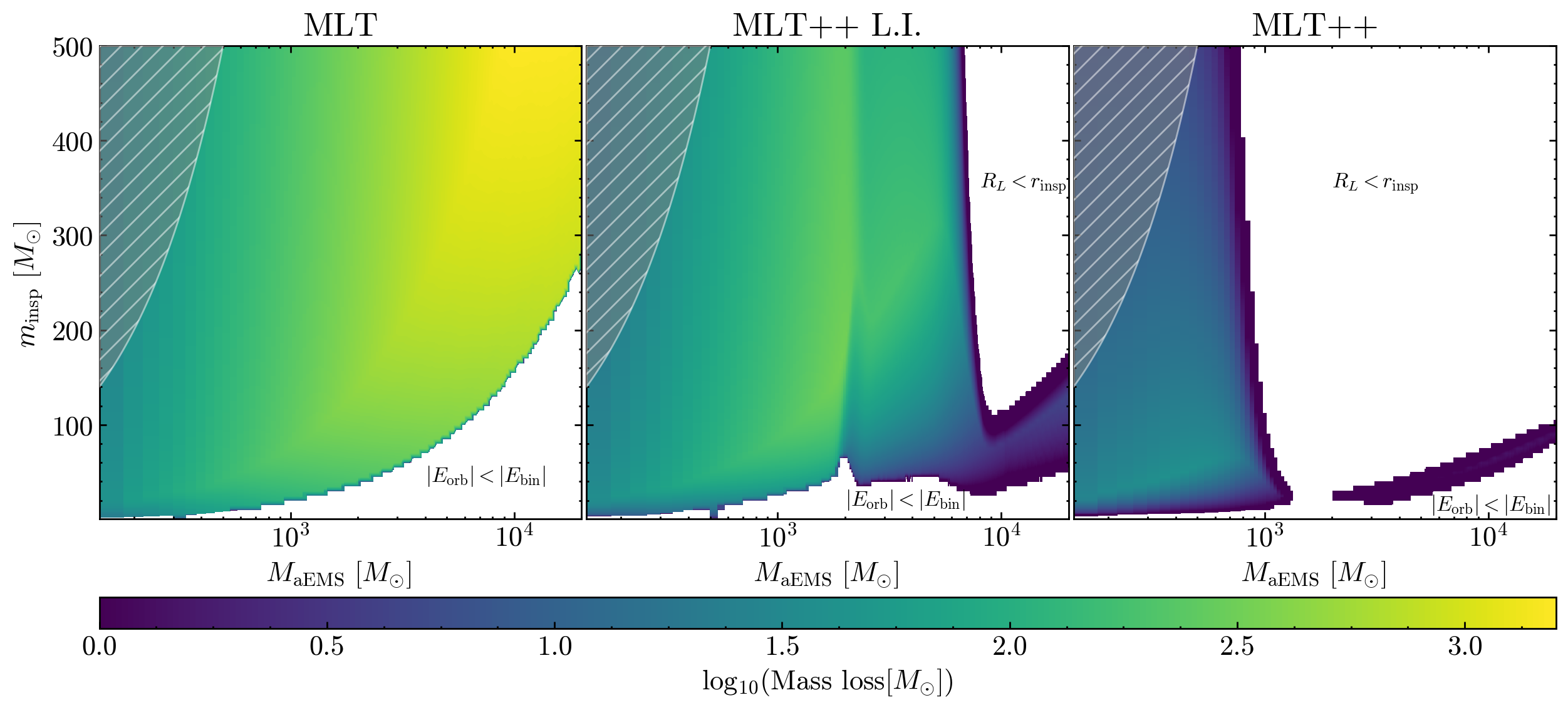}
    \includegraphics[width=\textwidth]{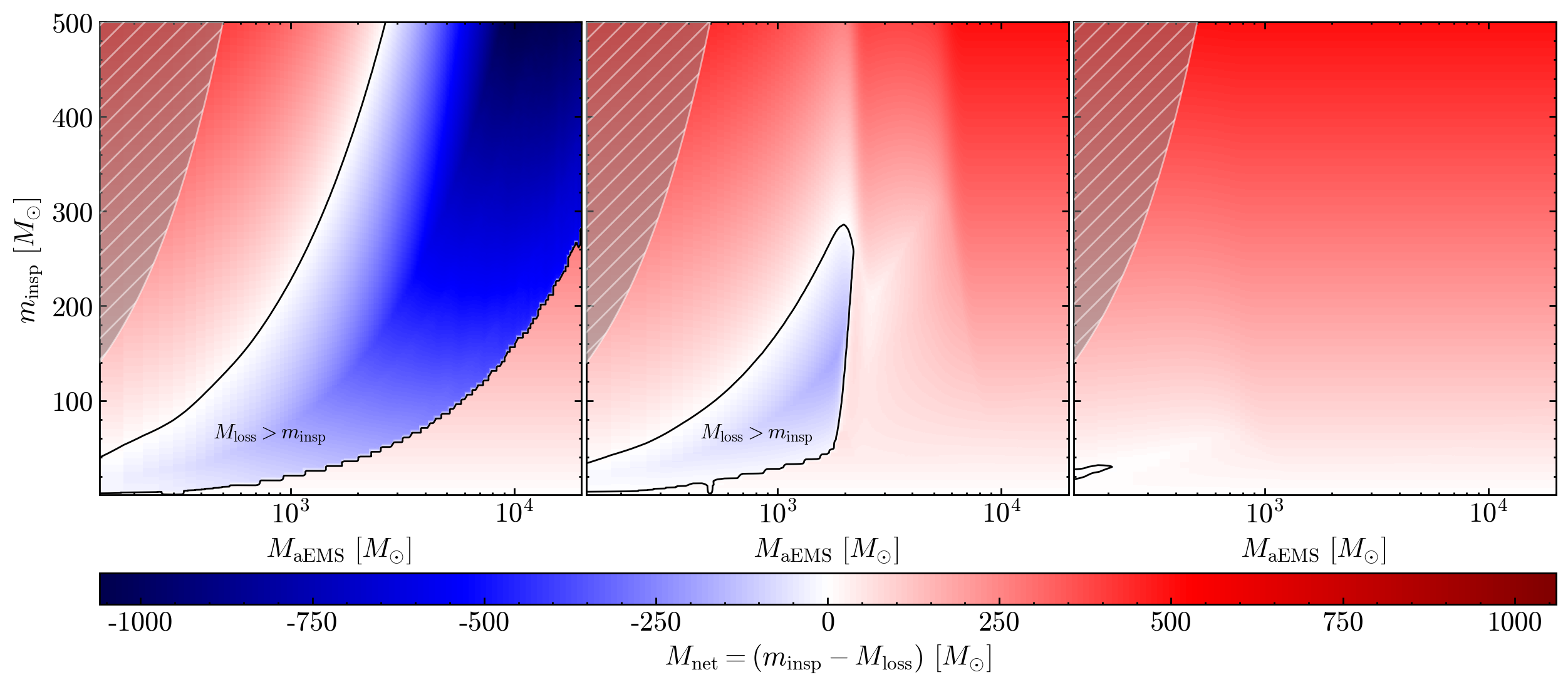}
    
    \caption{Color maps of the mass loss (upper row) and net mass change (lower row) due to a collision with inspiraling stars with $ m_{\mathrm{insp}}$ between 0.1 and 500 $M_{\odot}$ as a function of the mass of the target stars with $ M_\mathrm{aEMS}$  between 140 and 20.000 $M_{\odot}$. The three panels in each row correspond to the different convection treatments, as indicated. 
    The hatched area delineates the region where our analytical approach is not applicable, namely, $m_{\text{insp}}/M_\text{aEMS} \geq 1$. In the bottom row the black lines indicate the change in the sign of $M_{\rm net}$.}
    \label{fig:interpolation}
\end{figure*}

In Fig.~\ref{fig:energy_mltpp} we compare the absolute value of the binding energy $|E_\mathrm{bin}|$ within the interior of a $10^3 M_{\odot}$ aEMS model computed with MLT to the absolute value of the orbital energy released $|E_\mathrm{orb}^\mathrm{rel}|$ by a $30.0 M_{\odot}$ inspiraling star along its journey towards the aEMS's center (the model is for [Fe/H]=-2.00~dex). 
We indicate the depth at which the orbital energy released is higher than the binding energy of the outer layers of the aEMS, as well as the depth at which the merger can occur due to the condition of the local temperature ($T_\mathrm{aEMS} > T_\mathrm{vir}$) or because the inspiraling star fills its Roche lobe ($R_{L} < r_\mathrm{insp}$). 
The grey area indicates where these three conditions are met and shows the amount of mass that might be removed from the aEMS. For the specific case shown in Fig.~\ref{fig:energy_mltpp}, we expect that the mass loss caused by the collision is about $\sim 14\%$ of the mass of the aEMS, which corresponds to around $141.4 M_{\odot}$. This inspiraling star can effectively remove more than four times its own mass from the aEMS. 
Subsequently, we examine the mass loss anticipated under the aforementioned assumptions for a diverse range of inspiraling and target stars, and the influence of the mass-radius relations derived from the distinct approaches to convection. 

\subsection{Collision-induced mass loss and mass gain predictions for a large mass range of inspiraling and target stars}
\label{Analytical_predictions}

We applied the above analytical prescriptions and criteria to different combinations of masses for the inspiraling and the target stars assuming the value of [Fe/H]=-2.00~dex for their metallicity. 
We provide the predictions in tabular form for use by the community performing N-body and hydro-simulations of star clusters.
For this reason, we consider inspiraling stars with masses $m_{\text{insp}}$ between $0.1 M_{\odot}$ and $500 M_{\odot}$ 
and targets with masses $M_\text{aEMS}$ \footnote{In the following, we use $M_\text{aEMS}$  even if the mass range of the targets we consider includes VMS and SMS.}  ranging from $140M_{\odot}$ to $20000 M_{\odot}$.  Our analytical analysis is applicable only if  $m_{\text{insp}}/M_\text{aEMS} < 1$. 
Actual calculations were made with mass steps for the inspiraling stars of $0.1$ and  $5.0 M_{\odot}$ for masses between $0.1$ and $10M_{\odot}$ and between $10$ and $500M_{\odot}$ respectively. 
We derived the values of mass loss and net mass gain over  the entire domain by linear interpolation. 

The results are shown in Fig.~\ref{fig:interpolation} for the models with [Fe/H]=-2.00~dex and for the three different implementations for the transport of energy by convection. 
In the upper row, the colors scale with the amount of mass that can potentially be lost by the target of a given mass as a function of the mass of the inspiraling star. In the lower row, the colors show the corresponding net mass-change $M_{\rm net} = m_{\rm insp} - M_{\rm loss}$ for the target, with the blue and red regions corresponding respectively to a net mass loss ($M_{\rm net} < 0$) or mass gain ($M_{\rm net} > 0$).  

Starting the analysis with the MLT assumption, which produces more extended stars, we observe several key patterns in the $M_{\text{aEMS}}$-$m_{\text{insp}}$ plane.
The white area in the upper left panel of Fig.~\ref{fig:interpolation} corresponds to cases where no mass loss is expected because the orbital energy released by the inspiraling stars is insufficient to overcome the binding energy of the target aEMS. Consequently, the minimum mass of an  inspiraling star required to induce mass loss in the target aEMS increases with the target's mass. For an aEMS target of $1000 M_{\odot}$, the minimum inspiraling mass needed to produce mass loss is approximately $25 M_{\odot}$, resulting in a mass loss of about $126 M_{\odot}$ (i.e., $12.6\% $ of the mass of the target). For a target aEMS of $5000 M_{\odot}$, the minimum inspiraling mass needed to cause mass loss is around $90  M_{\odot}$, with a corresponding mass loss of approximately $375 M_{\odot}$ (i.e., $7.5\%$ of the mass of the target).   
There is a limited area in the middle part of the $ M_{\text{aEMS}}$-$m_{\text{insp}}$ plane where the amount of mass possibly lost by a target star exceeds 20$\%$ of its mass. Within this area, for example, an  inspiraling star of $100 M_{\odot}$ can induce a mass loss of 25$\%$ in a $1000 M_{\odot}$ aEMS. In the lower left panel, we see a large (blue) area of the parameter space where the collisions are expected to induce a net mass loss from the targets ($M_{\rm net} < 0$). This region is surrounded by two (red) areas where the target star actually increases its mass after the collision ($M_{\rm net} > 0$).

In the cases of MLT++ L.I. and MLT++, the aEMSs are more compact and have higher binding energy, making mass loss through collisions more difficult. As in the MLT case, in the upper panel, the white area located at the bottom part of the $M_{\text{aEMS}}$-$m_{\text{insp}}$ plane corresponds to cases where the collisions are not expected to induce mass loss, because the orbital energy of the inspiraling is insufficient to overcome the binding energy of the target aEMS. 
Furthermore, it is expected that no mass loss will occur in the white area located at the upper part of the plane. This is due to the Roche lobe criteria, which states that more compact aEMS stars result in smaller orbital distances in the orbit of the inspiraling star. Consequently, it is easier for the inspiraling stars to fill it according to Eq.~\ref{eq:roche_lobe}. In this case, we assume that the mass is transferred entirely to the aEMS and that the collision does not induce mass loss from the aEMS.  

In particular for the MLT++ L.I models, the radius of the aEMS model reaches a maximum when its total mass is around $1000M_{\odot}$, and it decreases later on (see panel (b) of Fig. \ref{fig:hrd}). At this stage, the minimum inspiraling mass required to induce mass loss is approximately $30M_{\odot}$, which can unbind about $5\%$ of the aEMS's mass.  Subsequently, as the aEMS becomes increasingly compact, it becomes more challenging for collisions to induce significant mass loss. For instance, for an aEMS with a mass of $2000 M_{\odot}$, the minimum inspiraling mass required to produce mass loss increases to approximately $70 M_{\odot}$. For aEMSs with masses between about $2500$ and $6000M_{\odot}$, inspiraling stars can induce a mass loss of up to $5\%$. However, for aEMSs with masses exceeding $6000M_{\odot}$, the mass loss cannot exceed $0.5\%$. 
As a result and as can be seen in the lower middle panel of Fig.~\ref{fig:interpolation}, the area where the target has a lower mass after the collision (i.e., $M_{\rm net} < 0$) is strongly reduced compare to the MLT case.

In the MLT++ case, the aEMSs are even more compact, resulting in higher binding energy. This compactness makes it more difficult for inspiraling stars to induce mass loss. This considerably reduces the region in the $M_{\text{aEMS}}$-$m_{\text{insp}}$ plane where collisions are expected to induce mass loss.  
Consequently, and as illustrated in the lower right panel of Fig.~\ref{fig:interpolation}, the mass injected by the inspiraling star is always higher than the mass loss it induces from the target whose mass always increases after the merger. 

This analysis underlines the necessity for a comprehensive treatment of the transport of energy by convection in aEMS that are radiation-pressure-dominated objects. However, our grid of models and the predictions for mass-loss and mass-gain induced by collisions over a large range of masses for both inspiraling and target stars provides for the first time an opportunity to explore how the stellar M-[Fe/H]-R-structure relation and its uncertainties impact the effects of collisions in the simulations of star cluster. Our grid can directly be  used in N-body and/or hydro simulations of star clusters.

\section{The impact of multiple collisions in a dense stellar cluster}
\label{multiplecollision}

\begin{figure*}
    \centering
    \includegraphics[width=0.95\textwidth]{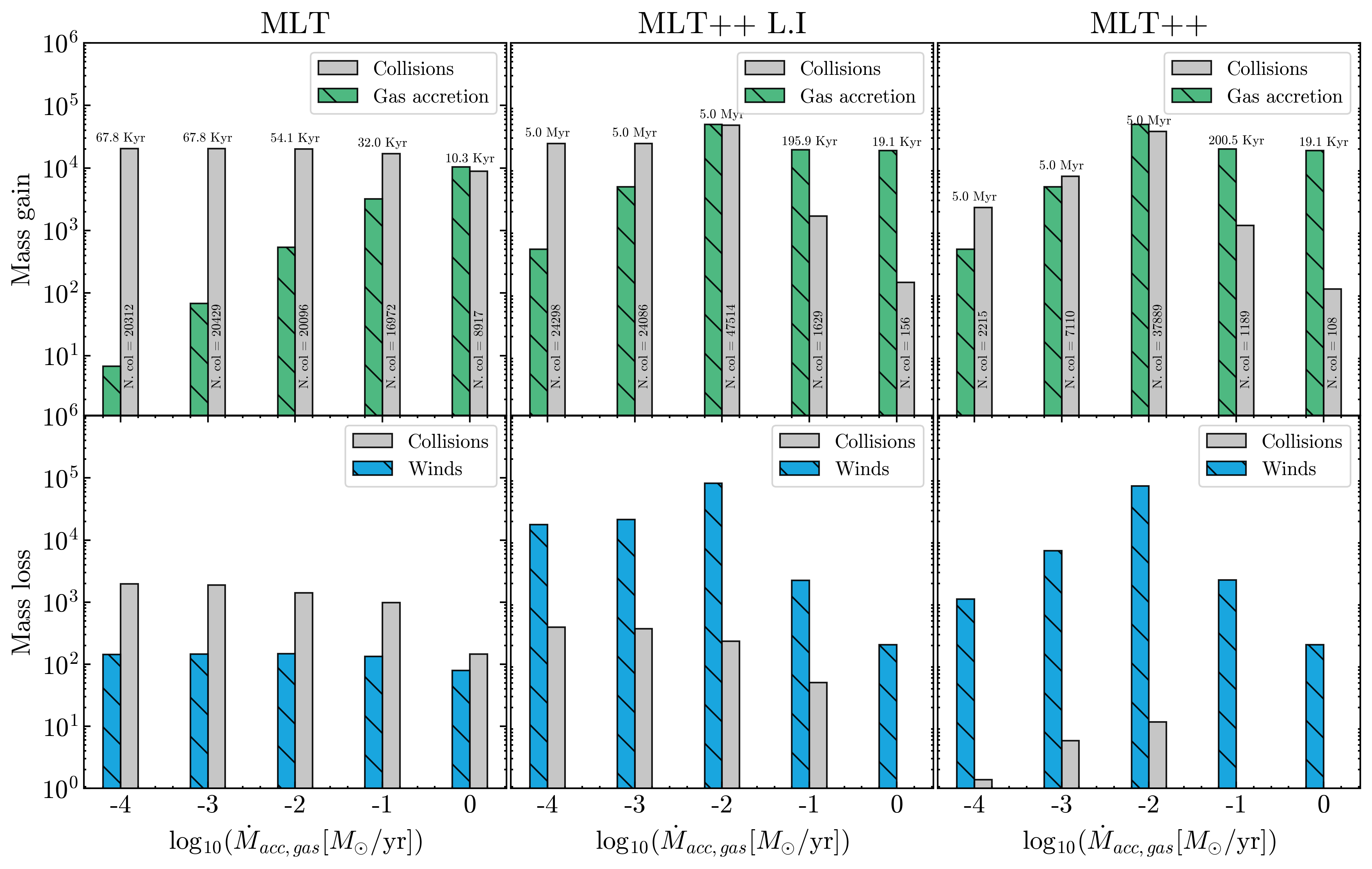}
    \caption{Top: Respective contributions of gas accretion and collisions to the total mass gain of the $1000\msun$ target for different gas accretion rates in the Monte Carlo simulations (mean values). Bottom:\ Respective contributions of winds and collision-induced mass loss to the total mass lost for the same simulations. The time above the bars shows the mean time to the end of the simulations, and the mean number of collisions is displayed in the corresponding bar chart.}
    \label{fig:MC3Myr}
\end{figure*}

In this section, we test our predictions for individual collisions in a simplified and illustrative way based on the  case of a static proto-GC cluster.
Our aim is to gain a deeper understanding of the circumstances under which collisions and mergers might lead to significant mass loss from the aEMS, as well as those where an aEMS in a typical proto-GC might enter into a conveyor belt regime where its mass gain is balanced by its mass loss. As in Sect.~\ref{onecollision},  here we focus on the models with [Fe/H]=-2.00~dex. 

\subsection{Model assumptions}

We followed the evolution of the mass of an aEMS target of initial mass of $1000M_{\odot}$ that undergoes multiple collisions with individual stars in a dense star cluster. 
We considered a static proto-globular cluster with a total initial number of stars of $10^5$ in the mass range between $0.1$ and $120$ $\mathrm{M_{\odot}}$. The masses of these stars were sampled according to the initial mass function (IMF) proposed by Kroupa (2001), leading to a total stellar mass in this mass range around $M_{0} \simeq 10^5 \mathrm{M_{\odot}}$.
We assumed a half-mass radius of $0.2$~pc, leading to a stellar density $\rho_\mathrm{h0}  \simeq 1.5 \times 10^6  \mathrm{M_{\odot}}/\mathrm{pc}^3$ and a stellar mass surface density of $\Sigma_h \simeq 7 \times 10^5 \mathrm{M_{\odot}}/\mathrm{pc}^2$.
This falls within the range of values derived for the candidate proto-GC at high redshift (Sect.~\ref{Introduction}).   

Without any preconceived hypothesis about its formation, we assumed that the aEMS is located at the center of the cluster,  it has the structure of the aEMS as predicted in Sect. \ref{Analytical_predictions}, and  it is the only target with which the stars can collide (we discuss the consequences of this latest assumption in Sect.
\ref{discussion}).

Following each collision, we modified the actual mass of the aEMS taking into account the expected collision-induced mass loss and gain predicted as described in Sect. \ref{Analytical_predictions} as well as the amount of mass the aEMS has lost by stellar winds and gained through accretion since the last collision (see also Sect. \ref{subsection:MC}).
For the mass loss due to the stellar winds, we followed the prescription from \citet{Vink2018}, namely, $\log \dot{M}_{winds} = -9.13 + 2.1 \log (M/M_{\odot}) + 0.74 \log (Z/Z_{\odot}) [M_{\odot} \mathrm{yr}^{-1}]$. Contrary to Sect. ~\ref{InputPhysics}, where the assumed mass-dependent high accretion rate to obtain the structure of the aEMSs, we now consider different values for the constant gas accretion rates during the collision phase,  namely, $\dot{M}_{acc,gas}= 10^{-4}, 10^{-3}, 10^{-2}, 10^{-1}$, and $1.0 \ M_{\odot} yr^{-1}$. As for the next collision, we consider the mass-radius relation and the structure predicted in Sect. \ref{Analytical_predictions} for the aEMS with the new mass calculated as described previously.
We derived the collision rate and incorporated it into a Monte Carlo-based framework, as described in Sections \ref{CollisionRate} and \ref{subsection:MC}, respectively.

\subsection{Collision rate in a static cluster}
\label{CollisionRate}

Following \citet{1976ApL....17...87H},  the collision rate of the target aEMS with cluster stars is given by 
\begin{equation}
    \dot{N}(M_{\text{aEMS}}, m_\text{insp}) = n_{\text {insp}} \gamma(M_{\text{aEMS}}, m_\text{insp})
,\end{equation}

where $n_{\text {insp}}$ is the average number of stars per unit of volume in the proto-globular cluster, and $\gamma$ is the rate coefficient with a star of specific mass $m_\text{insp}$ defined as 
\begin{equation}
  \gamma(M_{\text{aEMS}}, m_\text{insp}) = \frac{2\Sigma_{0}}{\sqrt{\pi l^{2}}}\left(1 + l^{2}v_{\infty}^{2} \right).
\end{equation}

Here, $\Sigma_{0}$ accounts for the geometric aspect of the effective cross-section,
\begin{equation}
    \Sigma_{0} = \pi(R_{\text{aEMS}}+r_\text{insp})^2 
    \label{crosssection}
,\end{equation}
and $v_{\infty}$ is the relative velocity required for two stars to close approach from infinity,
\begin{equation}
    v_\infty^2 = \frac{2G(M_{\text{aEMS}}+m_\text{insp})}{(R_{\text{aEMS}}+r_\text{insp})},
\end{equation}
with $R_{\text{aEMS}}$ and $r_\text{insp}$ their respective radii given from the mass-radius relation obtained from our accreting models. The parameter $l$ is inversely proportional to the dispersion of relative velocities. Assuming a static proto-globular cluster where all stars share the same mean square speed (i.e., an unsegregated cluster), we set 
\begin{equation}
    l^{2} = \frac{3}{4\langle v_{m}^{2} \rangle}, \text{with} \ \langle v_{m}^{2} \rangle \simeq 1.8 \frac{3GM}{8 R_{h}}. 
\end{equation}
From there, we can introduce the quantity $\tau(M_{\text{aEMS}}, m_\text{insp})$, which represents the average time in between collisions involving the aEMS with a star of mass $m_\text{insp}$. It is given by 
\begin{equation}
    \tau(M_{\text{aEMS}}, m_\text{insp}) = \frac{1}{\dot{N}(M_{\text{aEMS}}, m_\text{insp})}.
\end{equation}

\subsection{Monte Carlo-based simulations}
\label{subsection:MC}

The following algorithm was employed to manage collisions and their effect on the target aEMS over time. At a time \(t=0\) of a given simulation, the target aEMS star has a mass of $M_{\text{aEMS}}$. 
At each iteration, $i$, of the calculation, we followed the sequence detailed below: 
\begin{enumerate}
    \item A time step $\delta t$ of sufficiently small magnitude is selected to minimize the probability of multiple collisions, assuming $\delta t = \frac{\tau(M_{\text{aEMS}},m_{+})}{10}$,  where $m_{+}$ is the mass of the most massive protostar in the cluster. 
    \item A mass value $m_{k}$ is randomly selected from the IMF. 
    \item The probability of the collision between the target and the star of mass $m_{k}$ occurring during $\delta t$ is computed as  $p = \frac{\delta t}{\tau(M_{\text{aEMS}},m_{k})}$. 
    \item A random number, $u$, between 0 and 1 is drawn.
    \begin{enumerate}
        \item If $u < p$, we assume that the collision and the merger have occurred. 
        The target aEMS is assigned a new mass, given by 
        $M_\text{{aEMS}, i} = M_{\text{aEMS}, i-1} + m_{k} - M_{\text{col}, k} - M_{winds}  + M_{acc, gas}$ and the corresponding radius. 
        $M_{\text{col}, k}$ is the mass lost predicted in  \S~\ref{Analytical_predictions} for the collision
        obtained assuming the internal structure of the fast accreting model of mass  $M_\text{{aEMS}, i-1}$. Also, $M_{winds}$ and $M_{acc, gas}$ are, respectively, the mass lost by the winds and the mass gained from gas accretion by the aEMS during the time step $\delta t$. In the last step, a protostar of a specific mass, $m_{k}$, is removed from the initial sample of the IMF for the next iteration. 
        \item In the event where $u>p$, it is assumed that the aEMS is not subjected to a collision during the interval, $\delta t$, but still loses mass from winds and gains mass from gas accretion; the protostar of a specific mass, $m_{k}$, is kept in the IMF for the next iteration.
    \end{enumerate}

\end{enumerate}

The simulation is terminated when it reaches 5~Myrs. This is when the first supernova (SN),  stellar feedback, and binary heating effects are expected  to impact the cluster properties \citep[see e.g.,][] {Gieles+2018,Gieles+2024},  or when the mass of the aEMS falls below $150\mathrm{M_{\odot}}$ (lower limit) or rises above $2 \times 10^4 \mathrm{M_{\odot}}$ (upper limit). When the aEMS reaches the lower limit, it is assumed that the star will lose the majority of its mass. Conversely, when the star reaches the upper limit, it reaches a state where it becomes extremely challenging to cause further mass loss through collisions. In the MLT case, the minimum inspiraling star mass required to induce mass loss on a $2\times 10^4 M_{\odot}$ target is approximately $280M_{\odot}$, which we did not consider as we have assumed that the highest mass in the IMF is $120M_{\odot}$. We discuss this limitation further in Sect. \ref{Results-MC}.
Finally, given that this is an analytical approach, we do not account for the rejuvenation effects resulting from the collisions, nor the reaction of the star.

\subsection{Results of the Monte Carlo simulations}
\label{Results-MC}

We ran the Monte Carlo simulations 30 times for each combination of the following assumptions: 
transport of energy by convection (MLT, MLT++LI, and MLT++), and $\dot{M}_{acc,gas}$ over the ranges indicated above. 
We use the same initial mass distribution along the IMF for the stars for all the simulations. 
The results are presented in Fig.~\ref{fig:MC3Myr}. The x-axis indicates the assumed values for $\dot{M}_{acc,gas}$ (in $M_{\odot}/yr$). In the upper panels, the y-axis shows the respective mean values of the contributions of gas accretion and collisions to the total mass gain.  In the lower panels, it shows the respective mean values of the contributions of winds and collisions to the total mass loss.

A first examination of Fig.~\ref{fig:MC3Myr} indicates that the total quantity of material lost by the aEMS target and the respective contributions of the collisions and of the winds, are primarily influenced by the structural characteristics of the aEMS resulting from the treatment of energy transport by convection.
For a given treatment of convection, the second crucial parameter is the rate of gas accretion during the collision phase. When this quantity is varied across the three treatments of convection, it is found that there are configurations where the total mass lost by the aEMS target exceeds its initial mass. 

In the MLT case where the aEMS target has the most extended structure, its  effective cross-section is higher (see Eq.\ref{crosssection}), leading to the fastest rate of collisions for a given cluster configuration. Consequently, and even if the individual collisions are relatively efficient at inducing mass loss (see Fig.\ref{fig:interpolation}), the mass of the aEMS increases steeply thanks to the mergers. All the "MLT simulations" actually stop before $\sim$ 5Myrs, because the target star has reached the maximum mass of $2 \times 10^4 M_{\odot}$. 

The higher the gas accretion rate, the faster this happens (from $\sim 10$ to $\sim 68$ Kyrs for gas accretion rates of $1.0$ and $10^{-4}$ $ M_{\odot}.yr^{-1}$ respectively), with the mass gain of the aEMS being dominated by the mergers except for the highest value of the gas accretion rate (i.e., $1.0 M_{\odot}.yr^{-1}$), where the gas accretion starts to become dominant. Further calculations were conducted for $\dot{M}_{acc,gas}=0.01 M_{\odot}.yr^{-1}$, and an upper mass limit of $4 \times 10^4 M_{\odot}$, which was reached in $\sim 0.5$ Myr.  
Regarding the total mass loss, over these short timescales the collisions contribute more to the total ejecta than the stellar winds.
Notably, configurations in which the total mass lost exceeds the initial mass of the target occur predominantly in the low and intermediate gas accretion regimes, with the  exception of the case where $\dot{M}_{acc,gas}=1.0 M_{\odot}.yr^{-1}$.

In contrast, a more compact aEMS structure, obtained with MLT++L.I. and MLT++, results in a lower effective cross section for collisions. Furthermore, individual collisions are less efficient in inducing mass loss (Fig.\ref{fig:interpolation}). 

For the two highest gas  accretion rates considered ($\dot{M}_\mathrm{acc,gas} = 10^{-1}$ and $1.0 M_{\odot}/yr$), the aEMS reaches the upper mass limit in about  $19$ - $200$Kyrs. The total mass lost is approximately  $200$-$2300M_{\odot}$.

\begin{figure}
    \centering
    \includegraphics[width=\columnwidth]{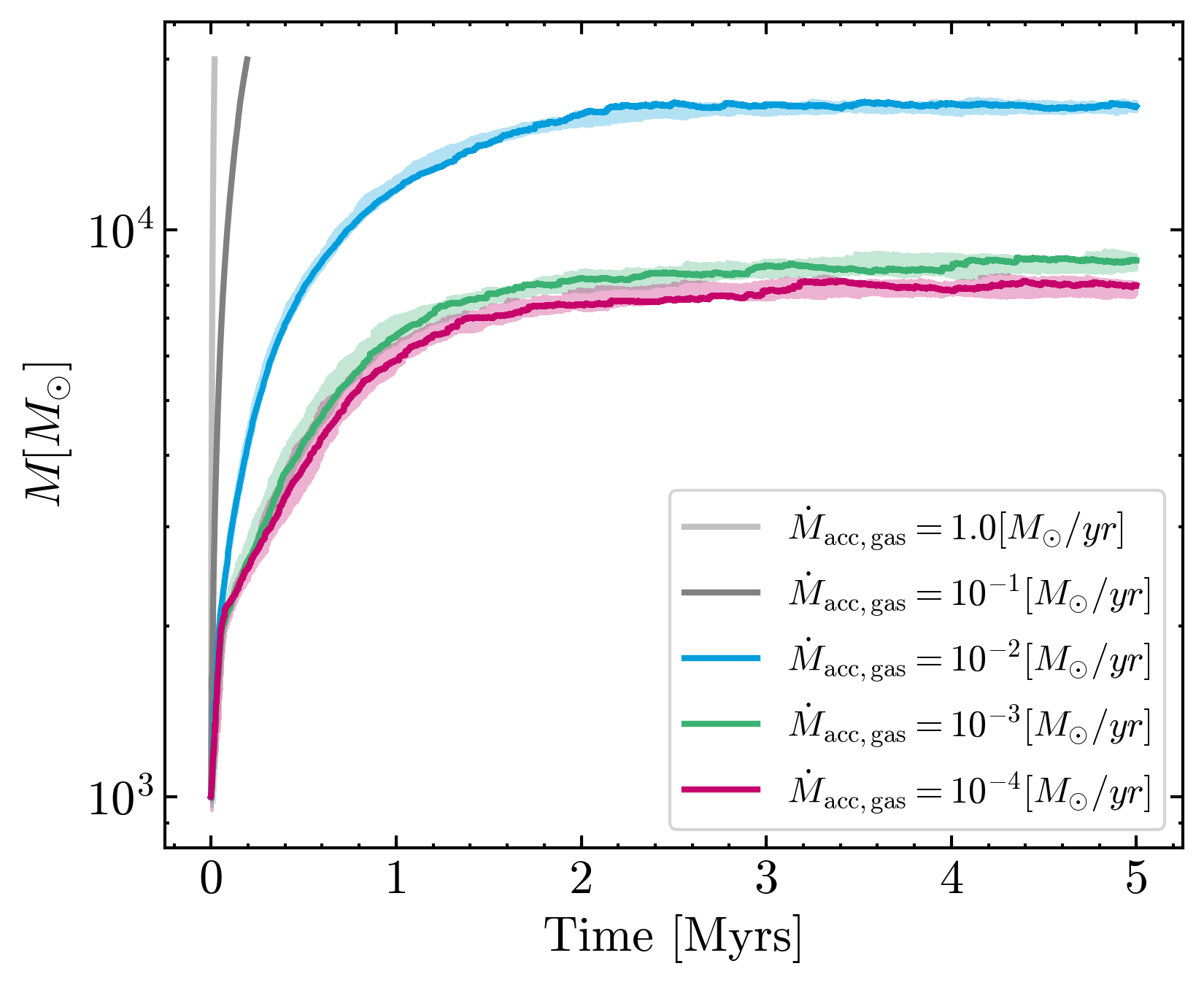}
    \caption{Trend of the evolution of the mass of an aEMS with an initial mass of $1000M_{\odot}$ ([Fe/H]=-2~dex) for the different accretion rates for MLT++L.I. The shaded regions enclose the 30 MC realizations, and the solid line shows one realization.}
    \label{fig:Evolution}
\end{figure}

On the other hand, all the MLT++L.I. and MLT++ simulations with the three lowest gas accretion rates exhibit a distinctive ``conveyor-belt'' behavior.
This is shown in Fig.~\ref{fig:Evolution} where we plot for the MLT++L.I. case the evolution of the mass of the target aEMS for the different $\dot{M}_{acc, gas}$. During the first $\sim 1 - 1.5$ Myrs, the mass of the aEMS target increases steadily, mainly due to the collisions and mergers with the stars, as shown in Fig.\ref{fig:MC3Myr}. 
After this initial mass-rising phase, the aEMS reaches a quasi-equilibrium state where the rates of accretion and mass loss balance each other. The equilibrium mass depends on the value of $\dot{M}_{acc, gas}$, a lower value of $\dot{M}_{acc, gas}$ resulting in a lower value of the equilibrium mass ($\sim 7500 - 8500 M_{\odot}$ for the two lowest values of $\dot{M}_{acc, gas}$, and $\sim 16000 M_{\odot}$ for the intermediate one, see Fig.\ref{fig:kipp} for the MLT+L.I. case). In these configurations, the corresponding evolution models are predicted to be fully convective. The chemical composition of the ejected material should then reflect that of the convective core, as expected to explain MPs in old GC and the peculiarities of the N-emitters. 
During the conveyor belt stage, the rate of mass gain (due to mergers and accretion of gas) and the rate of mass loss (due to collisions and winds) is approximately $\sim 4.45\times 10^{-3}$,  $5.35\times 10^{-3}$ and $2.08\times 10^{-2}$  M$_{\odot}/$yr, for $\dot{M}_{acc,gas}= 10^{-4}, 10^{-3} $and  $10^{-2}$ M$_{\odot}/$yr, respectively. The equilibrium is maintained until the end of the simulation.

This conveyor-belt behavior results in higher total mass loss compared to the results obtained with the MLT assumption. The maximum value for the total mass lost obtained over the range of parameters investigated, is approximately $4\times 10^{4} M_{\odot}$, for $\dot{M}_{acc,gas} = 10^{-2}$ M$_{\odot}/$yr within an interval of $5$Myrs.  Importantly, the amount of processed material liberated by the aEMS in the conveyor belt cases depicted in Fig.\ref{fig:Evolution} is  
one order of magnitude higher than its initial mass.

In the conveyor belt scenario, the aEMS is supposed to be continuously rejuvenated with fresh hydrogen that is both accreted and brought by the merging stars that have not yet burned H (a large fraction of them have a Kelvin-Helmholtz time longer than 10~Myrs). 
This continuous injection of unburnt material from gas accretion and stellar mergers is expected to effectively 
sustain a main sequence-like structure for a longer period, delaying the aEMS from evolving along the main sequence and entering advanced evolutionary phases(see also Appendix~\ref{centralnucleosynthesis}).
While our Monte Carlo simulations indicate that the conveyor belt regime is reachable within the simplistic framework of our static proto-star cluster model, exact numbers for the mass budget should be considered with  caution. One of the main caveats is the assumption that the aEMS is the only target for collisions, with more realistic models showing that VMS can be formed by gas accretion and collisions in compact clusters \citep[][and references therein]{Gieles+2024}. Additionally, recent N-body stellar collision studies have shown that the mass function of the colliding stars is more top-heavy than the Kroupa IMF due to mass segregation in the simulated star clusters \citep{2024MNRAS.531.3770R}. The next obvious step is to incorporate our model predictions for collision-induced mass loss and gain into realistic high-resolution hydrodynamical and N-body simulations of the formation of massive clusters and their host stellar populations (including remnant BHs). 

Another limitation of our analysis is the assumption of circular orbits for the inspiraling stars, while the collisions typically expected in a cluster occur along hyperbolic orbits. This assumption might lead to an underestimation of the velocities and energies involved in the collisions. Furthermore, in all three cases; MLT, MLT++, and MLT++ L.I, the timescale between collisions is shorter than the Kelvin-Helmholtz timescale. This implies that the aEMS does not have enough time to radiate away the energy gained from each collision before the next one occurs.
In the future, it  will  be necessary to evaluate
the dynamical response of the aEMS following a collision.

\section{Summary and conclusions}
\label{discussion}

Two main channels can contribute to the formation of EMS in non-metal free environments, namely, gas accretion and collisions, with expected co-lateral effects such as those our work aims to investigate for the first time.
We proposed and applied analytical prescriptions and criteria to derive the mass loss induced by individual collisions when an  inspiraling star spirals and deposits energy into the envelope of a target aEMS.  

We considered the internal structure properties of the targets that we obtained for fast accreting stellar models computed with the public stellar evolution code MESA, using different treatments of super-adiabatic convection.
While we cannot  claim this is how aEMS form,   
our choice for the gas accretion rate to build the preparatory aEMS models was driven by the necessity to reach  extreme mass before burning too much H, as required by the MSP abundance properties. The global properties we obtain for the aEMS as a function of stellar mass agree with those with similarly high accretion rates from previous, studies where the impact the gas accretion rate on the M-R relation has been covered. We show  that the predicted M-R and corresponding structures also  strongly depend on the treatment of super-adiabatic convection.
The more efficient this mechanism is assumed to be, the more compact and the bluer the model is at a given mass for a given gas accretion rate. This highlights the critical necessity for a comprehensive understanding of the transport of energy 
by convection or other means, including turbulence in radiative layers with high entropy gradient caused by gas accretion,  to reliably predict the evolution path of accreting aEMS in the HRD and the extent of their envelope and of their convective core.
The question of what happens once the aEMS stops being rejuvenated either through gas accretion and/or collisions and evolves in more advanced phases is out of the scope of the paper, since MSP show only very modest He enrichment and no sign of He-burning yields. Our analytical method can, however, be applied in the future to evaluate collision-induced mass loss on the structure of more evolved stars.

For the assumed mass- and luminosity-dependent gas accretion rates that we apply to reach the mass of the aEMS targets (later used to investigate collision-induced mass loss), the central temperature and density of the models are only modestly affected by the treatment of super-adiabaticity. These quantities depend mostly on the metallicity of the initial chemical composition of the stars, which is assumed to be similar to that of the gas of the proto-GC, and on the actual mass of the proto-aEMS. 
Consequently, more metal-poor conditions, as well as more massive aEMS, lead to more extreme abundance variations of the isotopes involved in the CNO cycle and the NeNa and MgAl chains in the convective core of the models. This is consistent with the extensions of the C-N, O-N, Na-O, and Mg-Al anticorrelations among MPs in old GC that are observed to depend primarily on the total mass and metallicity of the host cluster. The gas-accretion timescales of our models being much shorter than the nuclear burning timescales, the He-enrichment resulting from this hot H-burning nucleosynthesis is very limited, although slightly higher in the most metal-poor models, in agreement with photometric derivations of He abundance variations among MPs in old GC.

The analytical approach we have proposed to quantity collision-induced mass loss enabled us to investigate a wide range of masses for both the inspiralings and the targets (0.1 -- 500 M$_{\odot}$ and 140 -- 20000 M$_{\odot}$, respectively) with three different values of [Fe/H] covering the metallicity range of old Galactic GC. We probed the consequences of the different assumptions about superadiabaticity on the estimation of the mass loss induced by single collisions. 
In all cases, we found no disruptive collision configuration. As anticipated, the expected mass loss due to the deposition of orbital energy is higher in the case of the most extended targets that have less bound envelopes. It can reach $\sim 20 - 25 \%$ of the mass of the target star in a limited area of the $M_{\text{aEMS}}$-$m_{\text{insp}}$ plane,  with a minimum $m_{\text{insp}}$ to induce mass loss that increases with $M_{\text{aEMS}}$. Inducing mass loss from more compact targets, such as those obtained when assuming highly efficient energy transport through convection, proves to be more challenging for two principal reasons. Firstly, because of their higher bounding energy, and, secondly, because the smaller orbital separation reduces the  Roche lobe radius, which is rapidly filled by inspiraling stars across a significant portion of the $M_{\text{aEMS}}$-$m_{\text{insp}}$ plane. These results appear to be qualitatively reasonable.
 However, the individual collision analysis was conducted using pre-computed target structures, without accounting for the hydrodynamic response of the target to the deposition of energy that might also induce the expansion of the star as well a internal mixing beyond classical convection.

To gain a comprehensive picture of the dynamics of the collisions and accurately quantify their impact in terms of mass loss, it is now necessary to use hydrodynamic simulations that can track the reaction and the structural readjustment of the target aEMS to single and multiple collisions. 
It should be noted, however, that hydrodynamic simulations of stellar inspirals, even in 1D \citep[e.g.,][]{Fragos+2019} are computationally challenging and expensive. 
Our analytical  approach is thus very powerful as it covers a vast range of parameters and can provide guidance on the most relevant configurations for hydrodynamical investigation. 
Furthermore, this study provides predictions that can be used to investigate the impact of multiple collisions on a target aEMS put in simulations of a more realistic clustered environment. We have made them publicly available together with the mass-radius relations of our accreting star models computed with the three different assumptions for superadiabaticity. 

Finally, we have studied collision-induced mass loss and mass gain from an aEMS and quantified its contribution to the mass budget for forming MPs in a proto-GC. 
We used the predictions for single collisions in Monte Carlo simulations of multiple collisions onto an initially 1000 M$_{\odot}$ aEMS target located at the center of a compact and dense star cluster. We assumed the target is surrounded by stars with masses between 0.1 and 120 M$_{\odot}$ populating the Kroupa IMF. We chose a number of stars and a cluster radius that resulted in a stellar mass surface density comparable to that determined for candidate proto-GC at a high redshift. We ran 30 Monte Carlo simulations for each combination of the following parameters, applying a super-adiabaticity treatment, while the value of the constant gas accretion rates (between $10^{-4}$ and 1.0M$_{\odot}$/yr) during the collision phase. 
We could map the mass loss from collisions and winds versus the mass gain from mergers and accretion, as well as the timescales involved, over this range of parameters. 
The uncertainty about super-adiabaticity has again important consequences, together with the assumption about the gas accretion rate. Target stars with the largest radii are subject to  the highest collision rate and reach the upper limit of $2 \times 10^4$ M$_{\odot}$ that we have imposed in our simulations in a very short timescale ($10^4$ -- $6 \times 10^4$ yrs). The total mass lost by the target is limited, even though it can surpass the initial mass of the target in a specific area of the parameter domain we explored. On the other hand, a more compact target is submitted to a lower collision rate. 
In this case, when the gas accretion rate is low or moderate, the target attains a relatively high quasi-equilibrium mass through mergers (up to $16 \times 10^3$ M$_{\odot}$ in 1 or 2 Myrs), whereby the rates of accretion and mass loss nearly balance each other out. Within the most favourable range of parameters, 
up to $10^{4.9}$M$_{\odot}$ were ejected  in 5Myrs by the conveyor target. 
Provided that the target is kept at an almost constant level of rejuvenation through the regular supply of fresh hydrogen, which prolongates its lifetime on the main sequence, this is sufficient to address the mass budget issue for MPs in old GC. 
A caveat of our Monte Carlo simulations is that we did not address how the 1000 M$_{\odot}$ target was formed (i.e., we just assumed very high initial gas accretion). Furthermore, we  considered that collisions happen only with that specific object. However, all the member stars are potentially subject to collisions with their masses changing accordingly, which might modify the slope of the stellar mass function by favoring the formation of a VMS and of more than one EMS \citep[see][and references therein]{2024MNRAS.531.3770R}. Finally, we considered a static proto-GC, while the very early dynamical evolution of a proto-star cluster and the corresponding collision rate shall be impacted by gas accretion, mass segregation, stellar feedback, and binary heating.

Very recent Monte Carlo star cluster simulations addressed the impact of these parameters as well as that of the initial conditions (e.g., number of stars, cluster
compactness, stellar initial mass function, and primordial binary fraction) on the IMF and find collisional regimes that form VMS and EMS \citep{2024ApJ...969...29G,vergara2025rapidformationmassivestar}. The growth of VMSs in massive star clusters has also been predicted in recent N-body simulations and star-by-star hydrodynamical simulations, with important consequences on the formation paths of IMBHs  \citep{2024MNRAS.531.3770R,2025arXiv250418620L}. However, these state-of-the-art studies do not account for the proto-stellar phase, which is characterized by a high proto-stellar radius and a loose internal structure, and they assume that colliding stars
merge without mass loss, unless a CE event is triggered. 
The next important step is to include our model projections into sophisticated Monte Carlo and collisional N-body simulations,
to explore novel rejuvenation pathways for colliding proto-stars and conveyor belt solutions for the mass budget, as well as the resulting stellar and BH mass function, which will help to elucidate the MP phenomenon in GC.

\begin{acknowledgements}
   We thank the referee as well as N.Lahén and A.Rantala for their useful comments. The authors acknowledge  support by the Swiss National Science Foundation (Project 200020-192039 PI CC and Project CRSII5$\_$213497 PI TF). MG acknowledges  grants PID2021-125485NB-C22, CEX2019-000918-M funded by MCIN/AEI/10.13039/501100011033 (State Agency for Research of the Spanish Ministry of Science and Innovation) and SGR-2021-01069 grant (AGAUR).
\end{acknowledgements}

\section*{Data availability }
The files needed to reproduce the aEMS MESA models the predictions for the mass loss and mass gain are available at  \href{https://github.com/LauraRamirezGaleano/Mesa_Inlist_SMS.git}{this repository.}

% \begin{thebibliography}{}

% \end{thebibliography}

\bibliographystyle{aa} % style aa.bst
\bibliography{bib-refs}

\begin{appendix}
\section{Hydrogen-burning nucleosynthesis, mass, and metallicity effects in aEMS models}
\label{nucleosynthesis-appendix}

\subsection{Input physics}
\label{Input-physics}
To probe the metallicity effects on the aEMS internal structure and nucleosynthesis, we computed models with the same assumptions as in Sect.~\ref{InputPhysics} for three values of [Fe/H] (-0.72, -1.14, and -2.00~dex) that correspond to those of
three well-characterized Galactic GCs, NGC 104 (47Tuc), NGC 2808, and NGC 6397 respectively. For each metallicity, we run models for the three different treatments for convection as described in Sect.~\ref{Convection}. In all cases, we consider $\alpha-$element abundance enhancement of [$\alpha$/Fe] $=+0.3$~dex and adopt solar mixture from \citet{Asplund2009} for the other isotopes. The chemical composition of the accreted material is the same as that of the initial proto-stellar seed and it is deposited at the surface of the star.
We employ the nuclear reaction network \texttt{sagb\_NeNa\_MgAl} from MESA, which includes 29 species from neutrons to $^{27}$Al, and contains the pp-chains, CNO, NeNa, and MgAl cycles. 
The equation of state is the standard MESA combination of the SCVH \citep{Saumon+1995}, OPAL \citep{Rogers&Nayfonov2002}, HELM \citep{Timmes&Swesty00}, PC \citep{Potekhin&Chabrier2010}, FreeEOS \citep[]{FreeOS} and  Skye \citep{Jermyn+2021} equations of state \citep{Jermyn_2023}.
We used \texttt{eps\_grav} as the standard stellar structure energy equation form \citep[see section 5 of ][]{Jermyn_2023}. 
The opacity tables are taken from \citet{Iglesias+1993,Iglesias+1996} for the solar mixture of \citet{Asplund2009}. At high temperatures ($T \geq  10^8$K), Compton scattering opacities are computed using the prescription of \citet{Poutanen2017}. The conductive opacities are taken from \citet{Cassisi2007} with the correction of \citet{Blouin2020}.

\subsection{Nucleosynthesis}
\label{centralnucleosynthesis}

\begin{figure}[ht!]
    \centering
    \includegraphics[width=\hsize]{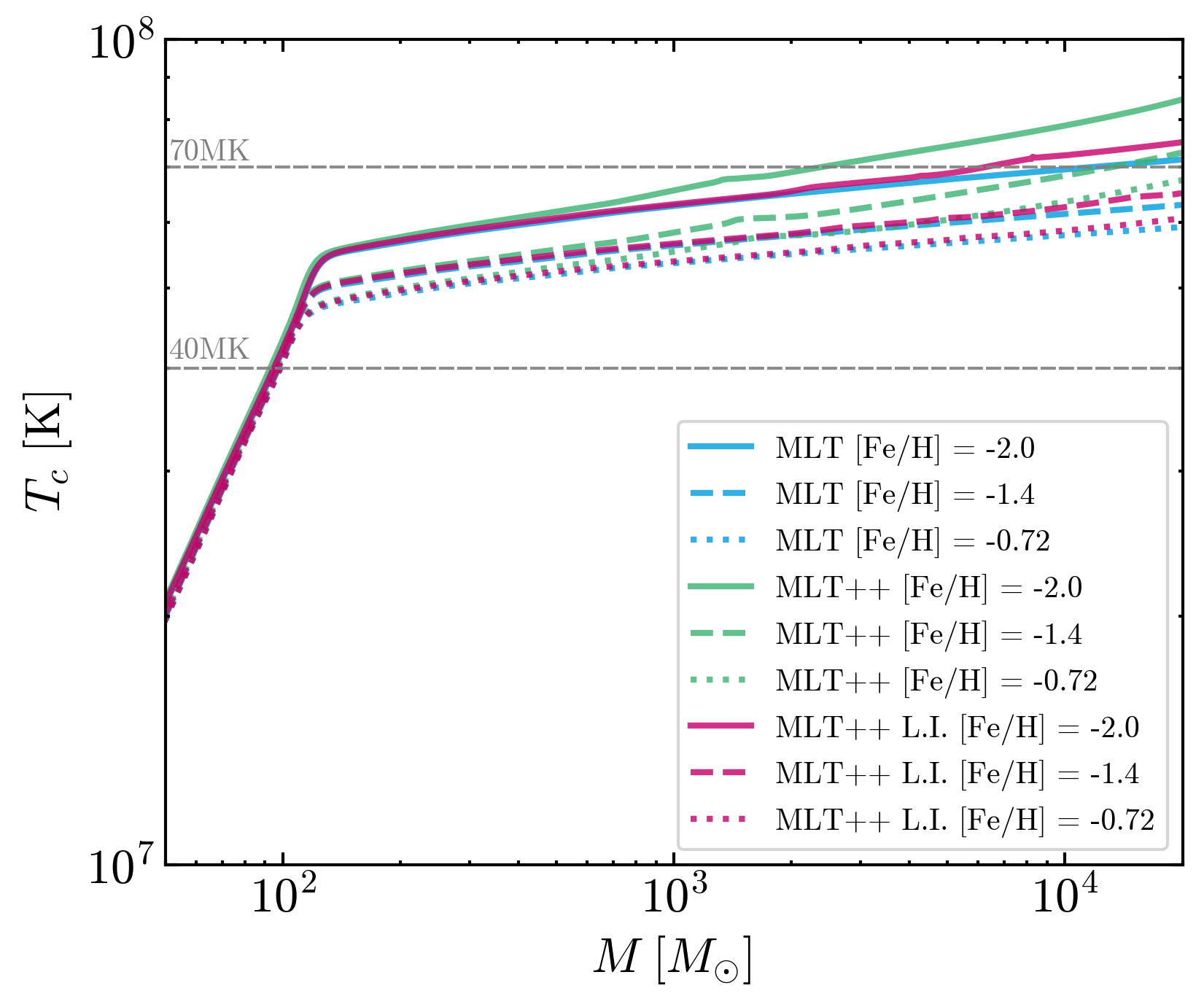}
    \caption{Central temperature as a function of the stellar mass for three different metallicities, [Fe/H] $= -0.72, -1.14,$ and $-2.00$ dex and for the three treatments for convection. 
    The horizontal dashed-gray lines are located at 40MK and 70MK respectively, as a reference.}
    \label{fig:T_c}
\end{figure}

\begin{figure}
    \centering
    \includegraphics[width=\hsize]{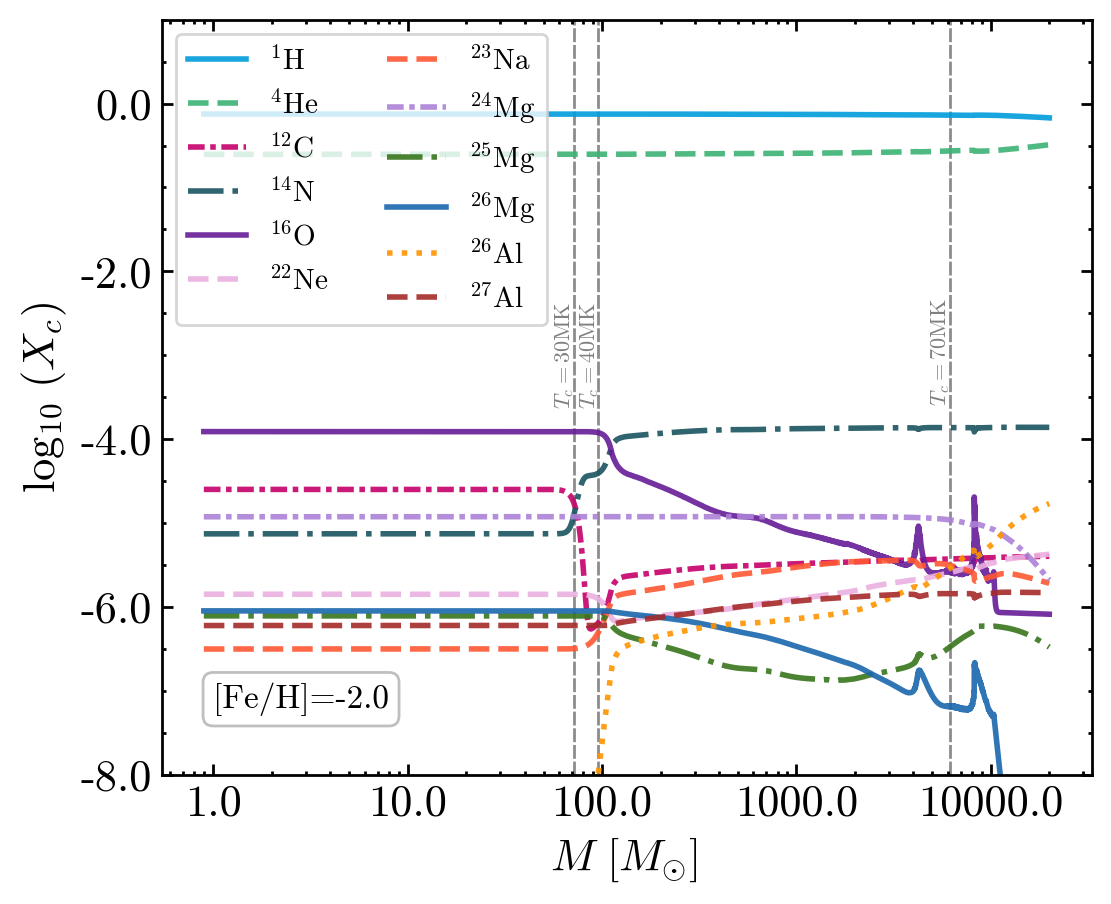}
    \includegraphics[width=\hsize]{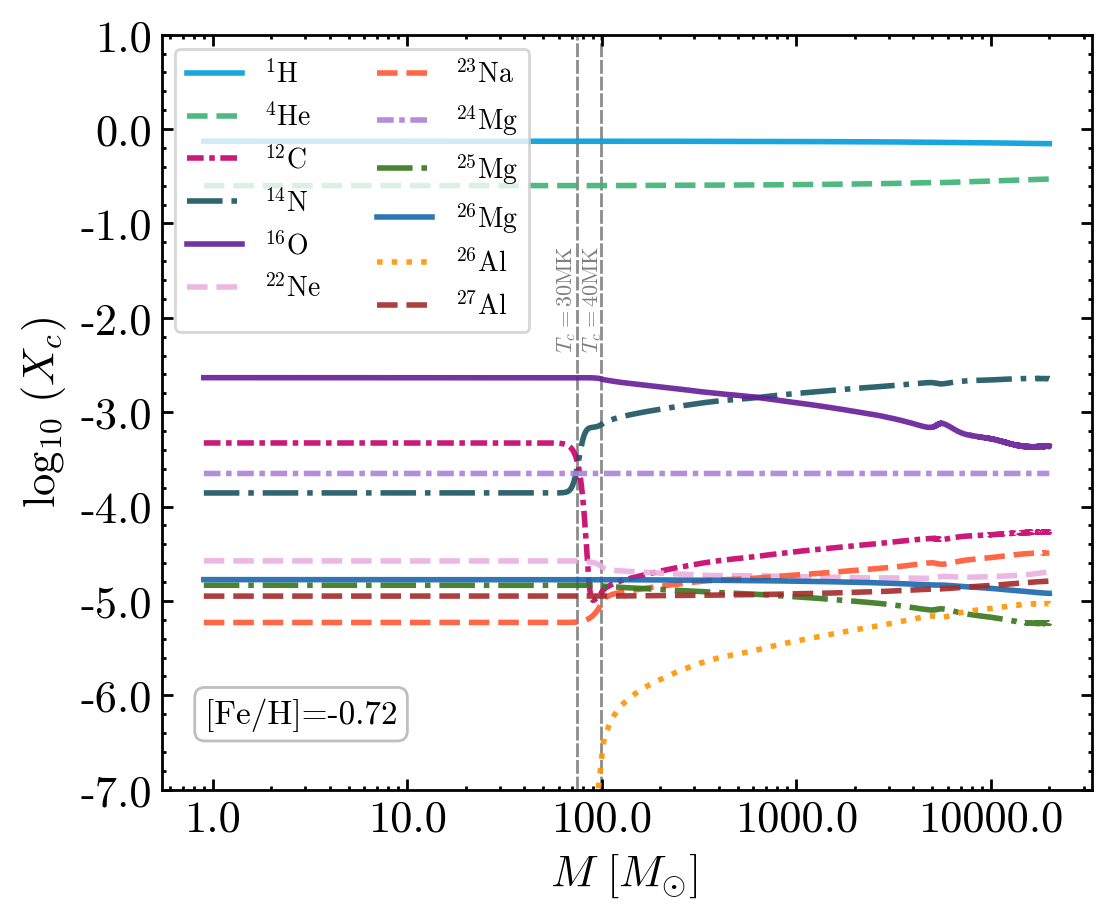}
    \caption{Evolution of the central abundances of relevant isotopes with respect to the mass of the star in the fast accreting models with [Fe/H]=-2 and -0.72~dex (top and bottom respectively; with MLT++L.I. in both cases). } \label{fig:central_abundances}
\end{figure}

Figure~\ref{fig:T_c} illustrates the relationship between the increasing central temperature and the stellar mass for all the convection and metallicity combinaisons. The curves initially overlap regardless of the [Fe/H] value and of the treatment of convection, until the stellar mass reaches approximately $10^2 M_{\odot}$. This is the point where the CNO-cycle reaches equilibrium at the very beginning of central H-burning, causing the core to expand and the central density to decrease (see also  Fig.~\ref{fig:hrd}). Later on, the more metal-poor models reach a higher central temperature at a given stellar mass, due to the impact of the metals on the stellar opacity. 
For each [Fe/H], the effect of the treatment of convection appears only when the stellar mass reaches $\sim$ 500-600$\msun$, and is more pronounced at lower metallicity. 
As a result, the central abundance variations of the isotopes involved in the CNO cycle, NeNa, and MgAl chains, as the stellar mass increases in the aEMS model, are more extreme in the model with [Fe/H]=$-2.0$~dex than in that with $-0.72$~dex, as shown in Fig.~\ref{fig:central_abundances}. 
Dashed gray lines mark the masses when the central temperature $T_c$ reaches specific values at which the abundances of specific isotopes start to change significantly in our models. 
For both metallicities,$^{12}$C starts to be heavily depleted while  $^{14}$N is increasing  around $T_c \sim 30$ MK. After the models have reached $\sim 10^2 M_{\odot}$ and the central temperature is above $\sim 40$~MK, the typical CNO equilibrium patterns build up (this is when core expansion occurs, as discussed above). $^{23}$Na and $^{26}$Al also start to increase (from $^{22}$Ne, and $^{25,26}$Mg respectively), while  $^{16}$O decreases significantly. Above $T_c \sim 60$ MK, namely, for masses above $\sim 10^3 M_{\odot}$ (few $10^3 M_{\odot}$) for the [Fe/H]=-2 (-0.74~dex) models, $^{23}$Na and $^{26}$Al have increased by more than $\sim$ 1.5~dex and almost 2~dex respectively.
The depletion of $^{24}$Mg results in more Al production for $T_c$  higher than $\sim 70$ MK.
For [Fe/H]=$-2.0$ ($-0.74$~dex) this is reached when the stellar mass is above $\sim 6 \times 10^3 M_{\odot}$ (few $10^5  M_{\odot}$\footnote{Not shown in Figure~\ref{fig:central_abundances}, as it the only model for which we pushed the computation to that mass.}).

Finally, when the stellar mass reaches $2 \times 10^4$~M$_{\odot}$, the helium enrichment is higher
in the most metal-poor model ($\Delta Y \simeq 0.07$ compared to 0.04), and $^{24}$Mg has decreased from $1.18 \times 10^{-5}$ to $2.09 \times 10^{-6}$). However, Mg depletion is not seen in most GC, but only in the most massive and/or metal-poor ones. This means that stars with masses above $\sim 10^4 M_{\odot}$ cannot have contributed in most cases (see \citealt{Gieles+2024} for more details about the Mg-Al anticorrelation).
Our results show that accreting models meet the GC nucleosynthesis constraints within a broad mass range, with the most massive and/or the most metal-poor stars leading to the largest abundance variations. The substantial mass content of the aEMS large convective core provides an optimal reservoir for addressing the mass budget issue.

It is important to note, however, that the exact relationship between stellar mass and abundance variations described above should be interpreted with caution. In addition to the influence of metallicity, the actual masses where similar abundance variations are reached shall also depend on the gas accretion rate. For slower accretion, He abundance variations would be higher at a given mass. Because the central temperature increases with the mean molecular weight, which would then increase in the core, the same abundance variations would be reached in slightly less massive stars than indicated above (see Fig.~4 in \citealt{Prantzos2017}). On the other hand, Na is expected to be depleted if the central temperature is too high. This will be quantified in greater detail in a subsequent study. However, this will not affect the conclusion of the present paper regarding collision-induced mass loss.

\end{appendix}

\end{document}